# Generalisation of continuous time random walk to anomalous diffusion MRI models with an age-related evaluation of human corpus callosum

Qianqian Yang [a, *] David C. Reutens [b,c] and Viktor Vegh [b,c]

[a] School of Mathematical Sciences, Faculty of Science, Queensland University of Technology, Brisbane 4000, Australia

[b] Centre for Advanced Imaging, University of Queensland, Brisbane 4072, Australia;

d.reutens@uq.edu.au (D.C.R.); v.vegh@uq.edu.au (V.V.)

[c] ARC Training Centre for Innovation in Biomedical Imaging Technology, Brisbane, 4072, Australia

* Corresponding author. q.yang@qut.edu.au; Tel: +61-7-3138-2890.

**ABSTRACT**

Diffusion MRI measures of the human brain provide key insight into microstructural variations across individuals and into the impact of central nervous system diseases and disorders. One approach to extract information from diffusion signals has been to use biologically relevant analytical models to link millimetre scale diffusion MRI measures with microscale influences. The other approach has been to represent diffusion as an anomalous transport process and infer microstructural information from the different anomalous diffusion equation parameters. In this study, we investigated how parameters of various anomalous diffusion models vary with age in the human brain white matter, particularly focusing on the corpus callosum. We first unified several established anomalous diffusion models (the super-diffusion, sub-diffusion, quasi-diffusion and fractional Bloch-Torrey models) under the continuous time random walk modelling framework. This unification allows a consistent parameter fitting strategy to be applied from which meaningful model parameter comparisons can be made. We then provided a novel way to derive the diffusional kurtosis imaging (DKI) model, which is shown to be a degree two approximation of the sub-diffusion model. This link between the DKI and sub-diffusion models led to a new robust technique for generating maps of kurtosis and diffusivity using the sub-diffusion parameters $\beta_{SUB}$ and $D_{SUB}$. Superior tissue contrast is achieved in kurtosis maps based on the sub-diffusion model. 7T diffusion weighted MRI data for 65 healthy participants in the age range 19-78 years was used in this study. Results revealed that anomalous diffusion model parameters $\alpha$ and $\beta$ have shown consistent positive correlation with age in the corpus callosum, indicating $\alpha$ and $\beta$ are sensitive to tissue microstructural changes in aging.

## 1 INTRODUCTION

Diffusion MRI (dMRI) provides a mechanism by which microscopic water diffusion in tissue can be probed. Diffusion-weighted MRI signals from tissue, such as the human brain, are obtained by varying the diffusion weighting in the dMRI protocol - primarily the duration, separation and strength of the diffusion gradient are changed to provide sensitisation to diffusion. To make a link between mm-scale measurements, and the micro-scale water interactions within tissue, mathematical models have widely been adopted to describe the expected diffusion signal behaviour. Through model parameter variations, *in vivo* information on tissue microstructure, such as bulk tissue diffusivity, anisotropy, structure symmetry and directional diffusivity (Basser et al., 1994), axon radius (Assaf et al., 2008), neurite orientation and density (Zhang et al., 2012) and level of diffusional kurtosis (Jensen et al., 2005) have been able to be inferred from dMRI measurements. Extensive applications of dMRI have been reported in the brain (Johansen-Berg and Behrens, 2014; Jones, 2010) and for other body parts (Brancato et al., 2019; Khoo et al., 2011; Miller et al., 2004; Taouli and Koh, 2009). The clinically routine use of dMRI in the clinic underpins the importance of this MRI contrast mechanism (González et al., 1999; Warach et al., 1995).

Whilst existing models have been derived based on the classical diffusion process with consideration given to the domain in which diffusion occurs, models capturing the anomalous nature of diffusion in tissue based on continuous time random walk (CTRW) theory (Klages et al., 2008; Metzler and Klafter, 2000) have been proposed as well. The primary difference between the two approaches lies in how the problem is considered. In the former, a physiological property is generally identified, and tissue geometry is incorporated into the model formulation. As such, a specific model parameter is assumed to relate directly to the property of interest, such as water compartments, axon diameter, neurite density, intra- and extra-cellular volume fractions (e.g. bi-exponential (Clark and Le Bihan, 2000), CHARMED (Assaf and Basser, 2005), AxCaliber (Assaf et al., 2008), ActiveAx (Alexander et al., 2010), NODDI (Zhang et al., 2012), VERDICT (Panagiotaki et al., 2015))). The second approach (i.e. CTRW) makes no microstructure assumptions but instead considers diffusion within a hindered and restricted complex micro-environment, wherein the mean-squared displacement of diffusion spins follows a power-law relaxation in time, $\langle x^2(t)\rangle \sim t^{\beta/\alpha}$, where $0.5 < \alpha \leq 1$ and $0 < \beta \leq 1$. The power $\beta/\alpha$ is the result of the competition between sub-diffusion and Lévy flights, where $\beta$ governs the long-tailed waiting time probability density function (pdf) in sub-diffusion processes, and $\alpha$ governs the long-tailed jump length pdf for Lévy flights (Metzler and Klafter, 2000). Hence, $\alpha$ and $\beta$ can be considered as a measure of the complexity of tissue microstructure.

The CTRW model has been recently applied in the context of dMRI (Gatto et al., 2019; Ingo et al., 2015, 2014; Karaman et al., 2016; Magin et al., 2014; Tang and Zhou, 2019; Yang et al., 2020; Yu et al., 2018; Zhong et al., 2019). Other anomalous diffusion models for dMRI include the super-diffusion (i.e., stretched exponential) (Bennett et al., 2003; Capuani et al., 2013; Hall and Barrick, 2008; Palombo et al., 2011), sub-diffusion (Bueno-Orovio et al., 2016; Capuani et al., 2013; Palombo et al., 2011), quasi-diffusion (Barrick et al., 2020), and fractional Bloch-Torrey (Magin et al., 2008) models.

Since these models were developed independently and fitted with different parameter fitting strategies, our first goal here is to unify existing anomalous diffusion models within the CTRW framework. This unification provides a foundation which allows for meaningful assessment on the tissue contrast provided by various parameters, and how well models are able to fit the data. In addition, diffusional kurtosis imaging (DKI) also aims to provide measurements of non-Gaussian diffusion in tissue (Jensen et al., 2005), so our second goal is to establish a mathematical link between DKI and models under the CTRW framework.

Many studies have shown that anomalous diffusion model parameters decrease as the microstructural complexity of tissue increases, such as in brain tumours (Barrick et al., 2020; Karaman et al., 2016; Sui et al., 2015; Yang et al., 2020), small vessel disease (Barrick et al., 2020), Parkinson disease (Zhong et al., 2019), ischemic stroke (Barrick et al., 2020; Grinberg et al., 2014), and liver fibrosis (Anderson et al., 2014). However, little research has been performed on investigating how anomalous diffusion model parameters vary in the white matter of aging brains, which are characterised by microstructural alterations, such as decreased axon density and increased axon radius. Since the corpus callosum is one of the most studied white matter structures in aging (Ardekani et al., 2007; Fan et al., 2019; Ota et al., 2006; Pfefferbaum et al., 2000; Pietrasik et al., 2020; Salat et al., 2005) and its microstructural heterogeneity is well documented (Aboitiz et al., 1996; Aboitiz et al., 1992), our third goal is to provide insight into how anomalous diffusion model parameters behave in corpus callosum with aging, and to promote the use of anomalous diffusion models in probing tissue microstructure.

## 2 THEORY

In this section, we describe the theory of continuous time random walk (CTRW) model and describe how commonly used anomalous diffusion models can be unified under the CTRW framework.

### 2.1 CTRW-MRI modelling framework

In the context of CTRW theory, the jump lengths and waiting times of diffusing particles (walkers) follow probability distributions with infinite variance in contrast to the finite variance for Gaussian distribution in the context of classical random walk (Metzler and Klafter, 2000). The motion of CTRW particles in heterogeneous media can be described by a time-space fractional diffusion equation:

$$\frac{\partial^\beta P(x,t)}{\partial t^\beta} = D_{\alpha,\beta} \frac{\partial^{2\alpha} P(x,t)}{\partial |x|^{2\alpha}}, \qquad 0 < \beta \leq 1, \qquad \frac{1}{2} < \alpha \leq 1, \tag{1}$$

where $P(x,t)$ is the density of the diffusing particles at location $x$ (in units of mm) at time $t$ (in units of s), $\frac{\partial^\beta}{\partial t^\beta}$ is the time fractional derivative of order $\beta$ ($0 < \beta \leq 1$) in the Caputo sense, $\frac{\partial^{2\alpha}}{\partial |x|^{2\alpha}}$ is the Riesz space fractional derivative of order $2\alpha$ ($\frac{1}{2} < \alpha \leq 1$), $D_{\alpha,\beta} = D_{1,2} \frac{\tau^{1-\beta}}{\mu^{2(1-\alpha)}}$ is the generalised anomalous diffusion coefficient with units of mm$^{2\alpha}$/s$^\beta$, $D_{1,2}$ is the diffusion coefficient in the tissue in units of mm$^2$/s, $\mu$ and $\tau$ are the constants for preserving units. The solution of the above fractional diffusion equation (1) in Fourier space is:

$$p(k,t) = E_\beta\left(-D_{\alpha,\beta}|k|^{2\alpha}t^\beta\right), \qquad (2)$$

where $E_\beta(z) = \sum_{n=0}^{\infty} \frac{z^n}{\Gamma(1+\beta n)}$ is the single-parameter Mittag-Leffler function, $\Gamma$ is the standard Gamma function and by definition $E_1(z) = \exp(z)$.

In the context of diffusion MRI, $k$ in Eq. (2) represents the q-space parameter $q = \frac{1}{2\pi}\gamma G\delta$, $t$ represents the effective diffusion time $\overline{\Delta} = \Delta - \delta/3$ and $p(k,t)$ represents the signal intensity $S(q,\overline{\Delta})$, and so Eq. (2) can be used to describe the diffusion signal decay as (Magin et al., 2020):

$$S(q,\overline{\Delta}) = S_0 E_\beta\left(-D_{\alpha,\beta}q^{2\alpha}\overline{\Delta}^\beta\right), \qquad (3)$$

where $\gamma, \delta, \Delta$ and $G$ are defined as the gyromagnetic ratio, diffusion pulse duration and separation between the pulses, and diffusion gradient pulse amplitude. Additionally, $S_0$ is the signal when $G = 0$. Defining $b = q^2\overline{\Delta}$ in units of s/mm$^2$, Eq. (3) can be expressed in terms of b-values (Magin et al., 2020):

$$S(b,\overline{\Delta}) = S_0 E_\beta\left(-D_{\alpha,\beta}b^\alpha\overline{\Delta}^{\beta-\alpha}\right) = S_0 E_\beta\left(-\left(bD_{app}\right)^\alpha\right), \qquad (4)$$

where

$$D_{app} = \left(D_{\alpha,\beta}\overline{\Delta}^{\beta-\alpha}\right)^{\frac{1}{\alpha}} = \left(D_{1,2}\frac{\tau^{1-\beta}}{\mu^{2(1-\alpha)}}\overline{\Delta}^{\beta-\alpha}\right)^{\frac{1}{\alpha}} \qquad (5)$$

is the apparent diffusivity in units of mm$^2$/s under the CTRW framework and $\tau$ in units of $s$ and $\mu$ in units of mm are additional parameters ensuring the standard units of $D_{app}$ are preserved. Note the CTRW model was derived under the condition of the short pulse approximation, which means $\delta \ll \Delta$. However, in practice, this assumption is routinely violated on clinical MRI systems due to technical and safety limitations of in vivo MRI, meaning that gradient pulses have finite duration usually in the range 20 ms $< \delta < \Delta <$ 70 ms (Barrick et al., 2021).

Models that can be unified under the CTRW framework include:

i) the mono-exponential model (MONO) when $\alpha = \beta = 1$, which is the solution to the classical Bloch-Torrey equation:

$$S = S_0 \exp(-bD_{MONO}), \quad D_{MONO} = D_{1,2}; \tag{6}$$

ii) the super-diffusion model (SUPER) when $\beta = 1$ and $\frac{1}{2} < \alpha \leq 1$, which is also known as the stretched-exponential model or the space-fractional diffusion model (Bennett et al., 2003; Capuani et al., 2013; Hall and Barrick, 2008; Palombo et al., 2011):

$$S = S_0 \exp(-(bD_{SUPER})^\alpha), \quad D_{SUPER} = \left(D_{1,2}\mu^{2(\alpha-1)}\overline{\Delta}^{1-\alpha}\right)^{\frac{1}{\alpha}}; \tag{7}$$

iii) the sub-diffusion model (SUB) when $\alpha = 1$ and $0 < \beta \leq 1$, which is also known as the time-fractional diffusion model (Bueno-Orovio et al., 2016; Capuani et al., 2013; Palombo et al., 2011; Yang et al., 2020):

$$S = S_0 E_\beta(-bD_{SUB}), \quad D_{SUB} = D_{1,2}\tau^{1-\beta}\overline{\Delta}^{\beta-1}; \tag{8}$$

iv) the quasi-diffusion model (QUASI) when $\alpha = \beta$ ($\frac{1}{2} < \alpha, \beta \leq 1$) (Barrick et al., 2020):

$$S = S_0 E_\beta\left(-(bD_{QUASI})^\beta\right), \quad D_{QUASI} = \left(D_{1,2}\tau^{1-\beta}\mu^{2(\beta-1)}\right)^{\frac{1}{\beta}}; \tag{9}$$

v) the full CTRW model when $0 < \beta \leq 1$ and $\frac{1}{2} < \alpha \leq 1$ (Ingo et al., 2015, 2014; Karaman et al., 2016; Yang et al., 2020; Yu et al., 2018):

$$S = S_0 E_\beta(-(bD_{CTRW})^\alpha), \quad D_{CTRW} = \left(D_{1,2}\tau^{1-\beta}\mu^{2(\alpha-1)}\overline{\Delta}^{\beta-\alpha}\right)^{\frac{1}{\alpha}}; \tag{10}$$

vi) the fractional Bloch-Torrey (FBT) model with $\frac{1}{2} < \alpha \leq 1$ (Magin et al., 2008):

$$S = S_0 \exp(-(bD_{FBT})^\alpha), \quad D_{FBT} = \left(D_{1,2}\mu^{2(\alpha-1)}\left(\Delta - \frac{2\alpha-1}{2\alpha+1}\delta\right)/\overline{\Delta}^\alpha\right)^{\frac{1}{\alpha}}. \tag{11}$$

Note that parameters $D_{MONO}, D_{SUPER}, D_{SUB}, D_{QUASI}, D_{CTRW}$ and $D_{FBT}$ are model specific apparent diffusivities in units of mm$^2$/s, and they are a function of the effective diffusion time $\overline{\Delta} = \Delta - \delta/3$ (except for $D_{MONO}$ and $D_{QUASI}$), refer to section 5.3 for further discussion on time-dependant diffusion. Parameter $D_{\alpha,\beta}$ in Eq. (1) has fractional-order units (mm$^{2\alpha}$/s$^{\beta}$) and is a constant diffusion coefficient, i.e., not time dependent. In addition, we note that the FBT model and SUPER models share similarities, both taking the stretched exponential form, with the difference being a different scaling in the apparent diffusivity.

## 2.2 Diffusion phase diagram

Parameters of the anomalous diffusion models introduced in the previous section do not have a straightforward interpretation when fitted to the diffusion-weighted MRI signal. For this reason, we provide an explanation based on the previously introduced diffusion phase diagram (Metzler and Klafter, 2000). The CTRW model assumes waiting time and jump length are governed by long-tailed probability density functions, thus leading to infinite characteristic waiting time and jump length variance. The mean-squared-displacement of random walk, $x$, follows a power law time dependence (Metzler and Klafter, 2000):

$$\langle x^2(t) \rangle \sim t^{\beta/\alpha}, \quad \frac{1}{2} < \alpha \leq 1, \quad 0 < \beta \leq 1. \tag{12}$$

Since we standardised the models, the value of $\alpha$ (space fractional index) is in the range (0.5, 1] and the value of $\beta$ (time fractional index) is in the range (0, 1]. Figure 1 illustrates the diffusion phase diagram in this standard parameter space, with depictions of how different models behave as a function of $\alpha$ and $\beta$. In the case of quasi-diffusion (the $\beta/\alpha = 1$ line), the mean-squared-displacement grows linearly with time, which is a key feature of standard Brownian motion. In both the case of SUB (the $\alpha = 1$ line) and SUPER (the $\beta = 1$ line), the mean-squared-displacement follows a power law in time, Eq. (12). Since the fractional Bloch-Torrey (FBT) model takes the stretched exponential form, it is considered to be confined to the same region as the super-diffusion model (SUPER) on the diffusion phase diagram. In the case of CTRW, the mean-squared-displacement can display sub-diffusive nature when $\beta/\alpha < 1$, super-diffusive nature when $\beta/\alpha > 1$, quasi-diffusive nature when $\beta/\alpha = 1$ and normal diffusion when $\beta = \alpha = 1$. It should be also pointed out that for the CTRW model, wherein $\beta$ and $\alpha$ are both free parameters, the power law governing the mean-squared displacement is not unique and may be obtained by different sets of $\beta$ and $\alpha$. For example, a value of 0.8 for $\beta/\alpha$ can be obtained by setting $\beta = 0.8$ and $\alpha = 1$, or any combination of $\beta$ and $\alpha$ which fall on the dashed line shown in Figure 1. This issue does not present in the SUPER-, SUB- and QUASI-diffusion models, since only one parameter needs to be obtained and the solution should fall uniquely on the $\beta = 1$ line, $\alpha = 1$ line or $\beta/\alpha = 1$ line. Hence, careful consideration should be made towards how uniquely $\beta$ and $\alpha$ are estimated based on the CTRW model.

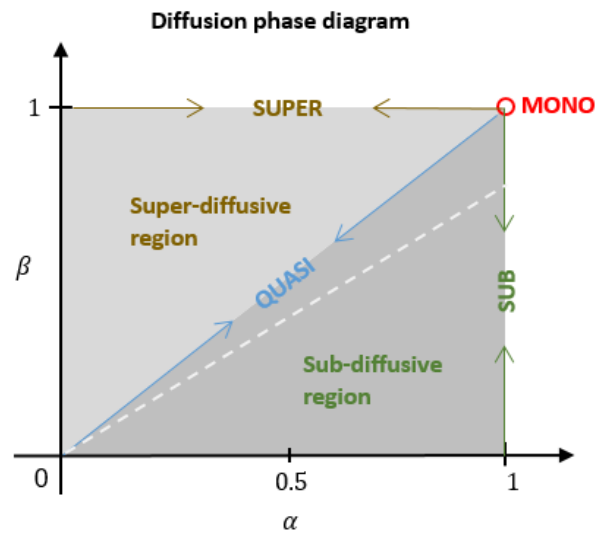

**Figure 1**. The diffusion phase diagram illustrated over a standardised domain described by $\alpha$ (space fractional index) and $\beta$ (time fractional index). Depicted are the action of the various models (MONO, SUPER, SUB, QUASI, and CTRW) within this domain. Note, CTRW can display sub-diffusive nature when $\beta/\alpha < 1$, super-diffusive nature when $\beta/\alpha > 1$, quasi-diffusive nature when $\beta/\alpha = 1$ or normal diffusion when $\beta = \alpha = 1$. The dashed line provides a hypothetical example when $\beta/\alpha = 0.8$, which is used to illustrate where the CTRW model may break down when diffusion is interpreted via mean-squared-displacement.

## 2.3 Diffusional Kurtosis Imaging

Diffusional kurtosis imaging (DKI) was developed in view of diffusion jump lengths not following standard Brownian motion (Jensen et al., 2005), hence jump lengths are assumed to deviate away from the Gaussian probability distribution function (pdf). The notion of kurtosis was used to measure the extent of deviation from Gaussian jump lengths, leading to:

$$S/S_0 = \exp(-bD_{DKI} + \frac{1}{6}b^2 D_{DKI}^2 K_{DKI}), \tag{13}$$

where $D_{DKI}$ is the apparent diffusion coefficient in units of $mm^2/s$, and $K_{DKI}$ is the parameter defining the level of kurtosis. To our knowledge, a direct link between the DKI formulation (13) and the sub-diffusion model (8) has not been made to date.

We will now show the DKI model (13) can be expressed as a degree two approximation of the sub-diffusion model (8). Since the parameters in the sub-diffusion model are real and positive, we are dealing with a real-value Mittag-Leffler function. This allows us to take the natural logarithm of the sub-diffusion model, Eq. (8), and obtain:

$$\log(S/S_0) = \log(E_\beta(-bD_{SUB})) = \log\left(\sum_{n=0}^{\infty} \frac{(-bD_{SUB})^n}{\Gamma(1+\beta n)}\right). \tag{14}$$

Taking the first three terms in (14) and applying Taylor series expansion at $b = 0$ gives:

$$\log\left(E_\beta(-bD_{SUB})\right) = \log\left(1 - \frac{bD_{SUB}}{\Gamma(1+\beta)} + \frac{b^2 D_{SUB}^2}{\Gamma(1+2\beta)} + O(b^3)\right)$$

$$= -\frac{bD_{SUB}}{\Gamma(1+\beta)} + \frac{b^2 D_{SUB}^2}{\Gamma(1+2\beta)} - \frac{1}{2}\left(-\frac{bD_{SUB}}{\Gamma(1+\beta)} + \frac{b^2 D_{SUB}^2}{\Gamma(1+2\beta)}\right)^2 + O(b^3)$$

$$= -\frac{bD_{SUB}}{\Gamma(1+\beta)} + \left(\frac{1}{\Gamma(1+2\beta)} - \frac{1}{2\Gamma^2(1+\beta)}\right) b^2 D_{SUB}^2 + O(b^3). \tag{15}$$

Letting

$$D^* = \frac{D_{SUB}}{\Gamma(1+\beta)} \tag{16}$$

and substituting $D_{SUB} = \Gamma(1+\beta)D^*$ into (15) gives

$$\log(E_\beta(-bD_{SUB})) = -bD^* + \left(\frac{\Gamma^2(1+\beta)}{\Gamma(1+2\beta)} - \frac{1}{2}\right) b^2 D^{*2} + O(b^3),$$

$$= -bD^* + \frac{1}{6} b^2 D^{*2} K^* + O(b^3), \tag{17}$$

with

$$K^* = 6\frac{\Gamma^2(1+\beta)}{\Gamma(1+2\beta)} - 3, \quad (18)$$

where $0 \leq K^* < 3$ for $0 < \beta \leq 1$. Note, when $\beta = 1$, the kurtosis $K^* = 0$, corresponding to the Gaussian case of diffusion. Note here the link between kurtosis $K^*$ and sub-diffusion parameter $\beta$, Eq.(18), coincides with the result in Ingo et al. (2015). The difference between Ingo et al.'s method and our method on deriving this result is provided in the Discussion, section 5.2.

According to Eq. (17), the DKI model can be written as a degree two approximation of the sub-diffusion model:

$$S/S_0 = E_\beta(-bD_{SUB}) \approx \exp\left(-bD^* + \frac{1}{6}b^2 D^{*2} K^*\right). \quad (19)$$

This link between the DKI model and the sub-diffusion model allows diffusivity $D^*$ and diffusional kurtosis $K^*$ to be computed from the sub-diffusion model parameters $D_{SUB}$ and $\beta_{SUB}$ according to Eqs. (16) and (18), respectively. The relationship between sub-diffusion model and its degree one to four approximations based on Eq. (17) is illustrated in Figure 2(a). Since DKI is taking quadratic form (yellow dashed line), to estimate the kurtosis based on DKI formulation, the valid range of b-value is limited to up to b = 3000 s/mm² (Jensen et al., 2005; Jensen and Helpern, 2010). But if using the sub-diffusion formulation, no such limitation on b-value is applied and in fact the kurtosis $K^*$ can be obtained as a complementary parameter based on $\beta_{SUB}$, see Eq.(18), with the relationship illustrated in Figure 2(b). Diffusivity, $D^*$, can be computed based on $D_{SUB}$ and $\beta_{SUB}$, see Eq. (16), with the relationship shown in Figure 2(c).

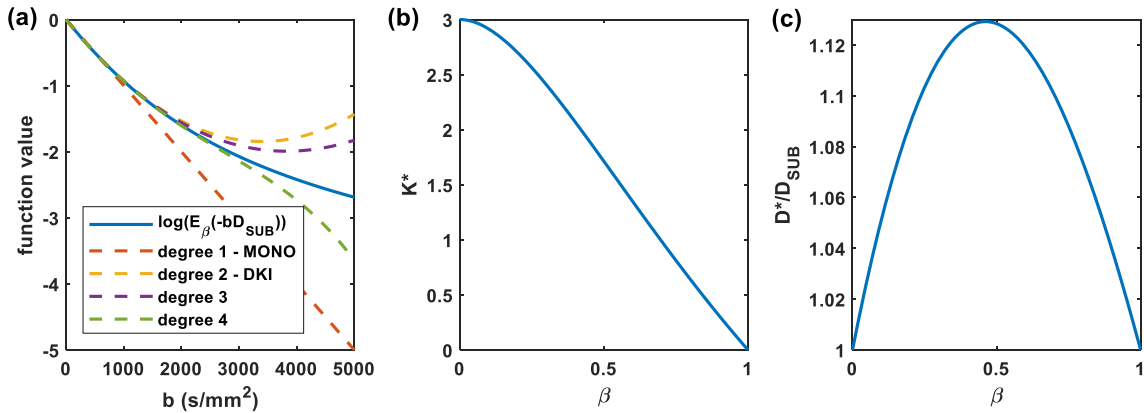

**Figure 2**. Link between sub-diffusion and DKI models: (a) plot of the natural logarithm of the sub-diffusion model and its degree one to four approximations according to Eq. (17). MONO and DKI are the degree one and two approximations of the sub-diffusion model, respectively; (b) relationship between kurtosis $K^*$ and $\beta_{SUB}$ according to Eq. (18); (c) relationship between the ratio $D^*/D_{SUB}$ and $\beta_{SUB}$ according to Eq. (16). $D^*$ and $K^*$ are the diffusivity and kurtosis computed from sub-diffusion model parameters $D_{SUB}$ and $\beta_{SUB}$.

## 3 METHODS

### 3.1 Simulation study

A key observation of white matter microstructure alteration related to normal aging is the change in axon radii (Stahon et al., 2016; Tang et al., 1997). To test if anomalous diffusion model parameters $\alpha$ and $\beta$ are sensitive to the microstructural changes in white matter with normal aging, we first performed a simulation study based on a white matter tissue model, where axonal fibres are represented as cylinders with a mean radius, R. We then evaluated how parameters mapped using the diffusion models Eqs. (6-11) vary with changes in axon radii.

This tissue model considers four signal compartments, $S_{ia}$, $S_{ea}$, $S_{csf}$ and $S_{tw}$, where *ia* is intra-axonal (restricted), *ea* is extra-axonal (hindered), *csf* is cerebrospinal fluid (free diffusion), and *tw* is trapped water in between myelin layers (stationary water). The model assumes water molecule exchange is at least an order of magnitude slower than diffusion in any of the four compartments, and therefore it is not modelled. The total diffusion MRI signal is given by

$$S^* = f_{ia}S_{ia} + f_{ea}S_{ea} + f_{csf}S_{csf} + f_{tw}S_{tw} \tag{20}$$

where $f$ represents volume fraction of each compartment with $f_{ia} + f_{ea} + f_{csf} + f_{tw} = 1$. The signal for each compartment is described as:

1. Intra-axonal restricted diffusion:

$$S_{ia} = S_0^* \exp(-bD_{ia}), \tag{21}$$

where $S_0^*$ is the MR signal with no diffusion weighting, $b = \gamma^2 G^2 \delta^2 \left(\Delta - \frac{\delta}{3}\right)$ is the diffusion weighting, $\gamma$ is the gyromagnetic ratio of water protons, $G$ is the amplitude of the pulsed gradients, and $\delta$ is the duration of the pulsed gradient. For a gradient with angle $\theta$ with respect to the axon axis, the effective intra-axonal diffusivity $D_{ia}$ is given by (Van Gelderen et al., 1994)

$$D_{ia} = \cos^2(\theta) D_{ia\|} + \sin^2(\theta) D_{ia\perp}. \quad (22)$$

Here $D_{ia\|}$ is the intra-axonal diffusivity along the axon axis, and $D_{ia\perp}$ is the intra-axonal diffusivity perpendicular to the axon axis, given by (Neuman, 1974; Van Gelderen et al., 1994)

$$D_{ia\perp} = \frac{2}{\delta^2 \left(\Delta - \frac{\delta}{3}\right)} \sum_{m=1}^{\infty} \frac{[2D_{ia\|}\alpha_m^2 \delta - 2 + 2\exp(-D_{ia\|}\alpha_m^2 \delta) + 2\exp(-D_{ia\|}\alpha_m^2 \Delta)) - \exp(-D_{ia\|}\alpha_m^2 (\Delta - \delta)) - \exp(-D_{ia\|}\alpha_m^2 (\Delta + \delta))]}{D_{ia\|}^2 \alpha_m^6 (R^2 \alpha_m^6 - 1)}, \quad (23)$$

where $R$ is the radius of the cylinder, and $\alpha_m$ are the roots of the equation

$$J_1'(\alpha_m R) = 0 \quad (24)$$

with $J_1'$ denoting the derivative of the Bessel function of the first kind, order one.

For long diffusion times and short gradient pulses, the signal perpendicular to the axon axis approaches (Van Gelderen et al., 1994)

$$S_{ia\perp} = S_0^* \exp(-\gamma^2 G^2 \delta^2 R^2), \quad (25)$$

meaning $D_{ia\perp}$ in Eq. (23) is reduced to

$$D_{ia\perp} = R^2 / \left(\Delta - \frac{\delta}{3}\right). \quad (26)$$

Note that Eq.(25) under the short pulse approximation was used in the ActiveAX white matter model (Alexander et al., 2010) to represent the intra-axonal diffusion signal. As pointed out by (Barrick et al., 2021), the short pulse approximation is routinely violated on clinical MRI systems due to technical and safety limitations of in vivo MRI. Hence, here we use the full model (21)-(23) to represent the intra-axonal diffusion signal.

2. Extra-axonal hindered diffusion:

$$S_{ea} = S_0^* \exp(-bD_{ea}), \quad (26)$$

where $D_{ea}$ is the effective extra-axonal diffusivity, given by

$$D_{ea} = \cos^2(\theta)D_{ea\parallel} + \sin^2(\theta)D_{ea\perp}. \tag{27}$$

The extra-axonal parallel diffusivity $D_{ea\parallel}$ is the same as the intra-axonal parallel diffusivity $D_{ia\parallel}$. The relationship between $D_{ea\parallel}$ and $D_{ea\perp}$ was set by a simple tortuosity model (Szafer et al., 1995) as

$$D_{ea\perp} = D_{ea\parallel} f_{ea}/(f_{ia} + f_{ea}), \tag{28}$$

assuming CSF and stationary compartments do not contribute to $S_{ea}$.

3. Free diffusion in CSF:

$$S_{csf} = S_0^* \exp(-bD_{csf}), \tag{29}$$

which is governed by isotropic Gaussian displacements.

4. Trapped water in subcellular structures such as the gaps between myelin lipid bilayers:

$$S_{tw} = S_0^* \tag{30}$$

which is not attenuated by the diffusion weighting.

To generate simulated DWI data, we set parameters in the tissue model as $D_{ia\parallel} = D_{ea\parallel} = 1.7\ \mu m^2/ms$ (Alexander et al., 2010), $D_{csf} = 3.0\ \mu m^2/ms$, $f_{ia} = 0.40$, $f_{ea} = 0.50$, $f_{csf} = 0$, $f_{tw} = 0.10$, $S_0^* = 1000$, $\Delta = 31.9$ ms, $\delta = 21.6$ ms, number of gradient directions = 42, and $b = 0, 0.5, 1.5, 2.5, 3.5$ ms/$\mu m^2$. Trace-weighted signal is then computed by taking the geometric mean of the generated multi-directional signals for each b-value.

Our choice of $f_{ia}$ and $f_{tw}$ resulted in a g-ratio = $\frac{R_i}{R_o} = \sqrt{\left(1 + \frac{0.85*f_{tw}}{0.4*f_{ia}}\right)^{-1}} = 0.8$ (Thapaliya et al., 2018) where $R_i$ and $R_o$ and inner and outer axon radii, and this value agrees with *in vivo* human corpus callosum g-ratio estimation (Berman et al., 2018; Jung et al., 2018). Note that the g-ratio has been shown to be consistent across corpus callosum sub-regions (Thapaliya et al., 2018) and shows little change with age (Berman et al., 2018). We should also point out that it has been shown that an increase in $R$ does not impact $f_{ia}$ and $f_{ea}$ considerably (Kodiweera et al., 2016). We varied $R$ in $S_{ia}$ and investigated how well diffusion models (6-11) were able to fit the simulated data and how diffusion parameters $(D_{app}, \alpha, \beta)$ for each diffusion model change as $R$ varies.

## 3.2 In vivo human diffusion-weighted MRI data

The study was approved by the human ethics committee of the University of Queensland, Brisbane, Australia. We recruited 65 healthy participants (30 females and 35 males) aged 19 to 78 years. The age distribution of the participants is presented in Figure S1 in the supplementary material. The dMRI data were collected using a 7T Siemens Magnetom research MRI scanner (maximum gradient strength G = 80 mT/m) with the Diffusion-Weighted Gradient Echo-Echo Planar Imaging (DW GE-EPI) sequence.

Diffusion experiments were performed with the following acquisition parameters: TE = 73 ms, TR = 7244 ms, δ = 21.6 ms, Δ = 31.9 ms, bandwidth = 1148 Hz/Px, isotropic resolution of 1.6 mm$^3$, matrix size = 132 × 132, number of axial slices = 70, field of view = 210 mm × 210 mm × 112 mm. By varying the gradient strength G, four different *b*-values were acquired: *b* = 500 (12), 1500 (24), 2500 (36), 3500 (48) s/mm$^2$. The number of diffusion directions (in parenthesis) were increased with increasing *b*-value. A total of 126 acquisitions including six *b* = 0 measurements were acquired for each participant; gradient directions at each *b*-value were chosen based on the electrostatic model (Jones et al., 1999; Landman et al., 2007). Acquisition of diffusion weighted MRI data took 16 minutes for each participant.

All diffusion-weighted MRI data were pre-processed in MRtrix3 (Tournier et al., 2019). The command *dwidenoise* (Cordero-Grande et al., 2019) was first used to denoise the diffusion MRI data. And then the command *dwifslpreproc* was used to call the eddy tool (Andersson and Sotiropoulos, 2016) in FSL (Smith et al., 2004) to correct for motion and eddy current distortions.

To remove the effect of diffusion anisotropy and to increase SNR, geometrically-averaged diffusion-weighted images (i.e. trace-weighted images) were computed across applied gradient directions for each *b*-value, and an example set of images is illustrated in Figure 3.

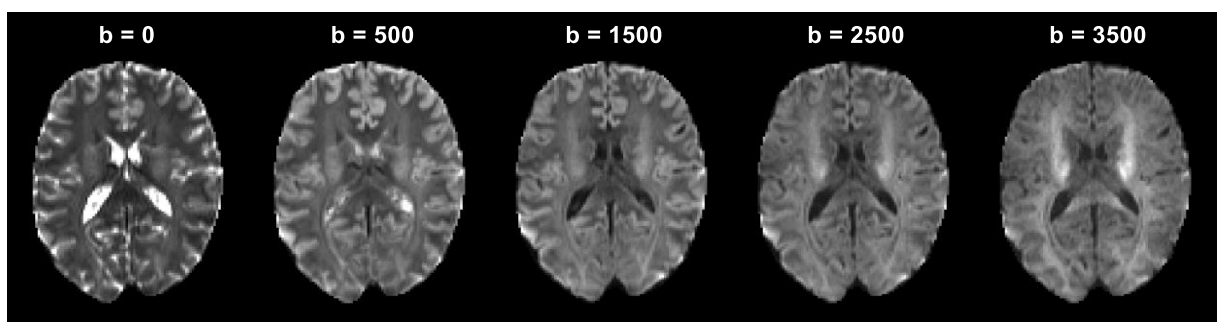

**Figure 3**. An example set of the trace-weighted axial slice images at different *b*-values (in units of s/mm$^2$) for a young healthy male participant.

### 3.3 Corpus callosum segmentation

In each human brain the corpus callosum was segmented manually into seven sub-regions (1 – rostrum, 2 – genu, 3 – rostral body, 4 – anterior midbody, 5 – posterior midbody, 6 – isthmus, 7 – splenium, illustrated in Figure 4) based on a standardised template (Witelson, 1989) and by using the MIPAV software (McAuliffe et al., 2001). In particular, to reduce the likeliness of partial volume effects, we drew regions-of-interest (ROIs) around one voxel inside the edge of the corpus callosum. Based on experience associated with our previous study (Yu et al., 2017), we found tractography-based segmentation of the corpus callosum to be consistent with the template-based method. Hence, here we adopted the template-based method to remove the need to perform seed-based tractography. The average number of voxels in the mid-sagittal section of corpus callosum was 556 ± 96. The average number of voxels in each subregion were 19 ± 9 in rostrum, 102 ± 29 in genu, 80 ± 25 in rostral body, 60 ± 14 in anterior midbody, 58 ± 15 in posterior midbody, 66 ± 17 in isthmus, and 199 ± 53 in splenium.

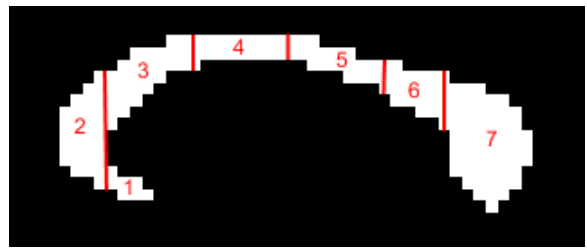

**Figure 4.** Diagram of the corpus callosum outlined around the mid-sagittal plane of a participant, depicting the seven subregions: 1 – rostrum, 2 – genu, 3 – rostral body, 4 – anterior midbody, 5 – posterior midbody, 6 – isthmus, 7 – splenium, according to the template in Witelson (1989).

### 3.4 Parameter estimation

Data smoothing was not performed prior to estimating anomalous diffusion model parameters. Each of the models in Eqs. (6-11) and Eq. (13) were fitted to the trace-weighted diffusion data in a voxel-by-voxel manner using MATLAB's *lsqcurvefit* function with the *trust-region reflective* algorithm. For the mono-exponential model, Eq. (6), $D_{MONO}$ was estimated using data with b = 0, 500, 1500 s/mm$^2$. For each anomalous diffusion model Eqs. (7-11), the fitted parameters are the corresponding apparent diffusivities, $D_{SUPER}$, $D_{SUB}$, $D_{QUASI}$, $D_{CTRW}$ or $D_{FBT}$, and the fractional indices $\alpha$ and/or $\beta$. To avoid overfitting, $D_{1,2}$, $\tau$ and $\mu$ were not estimated separately. Space fractional index $\alpha$ and time fractional index $\beta$ were bound in the range $(0.5, 1]$ and $(0, 1]$, respectively. Fitted model parameters were found not to be sensitive to the choice of initial values. Optimisation parameters *TolFun* and *TolX* were set to 1e-6 and 1e-4, respectively.

Because of the quadratic form of the DKI formula in Eq. (13), as elucidated in Figure 2(a), we used data with b = 0, 500, 1500, 2500 s/mm² for estimating $D_{DKI}$ and $K_{DKI}$.

### 3.5 Statistical analysis

Using the binary masks created in MIPAV, parameter values for each sub-region of the corpus callosum were extracted and averaged for that sub-region. The MATLAB functions *fitlm* and *predict* were used to perform linear and quadratic regression of ROI mean parameter values as a function of age and to find the 95% confidence interval. F-test was performed to determine the significance of the $Age$ and $Age^2$ terms. The significant level was set at $p \leq 0.05$. The Bonferroni method was used to correct for multiple comparisons between the seven subregions, i.e. uncorrected p-value needed to be less than 0.05/7 to be considered statistically significant in this study.

Following Barrick et al. (2020), we computed the tissue contrast as $TC = |\mu_{WM} - \mu_{GM}|/\sqrt{\sigma_{WM}^2 + \sigma_{GM}^2}$, where $\mu_{WM}$ and $\mu_{GM}$ are the mean parameter values in white and grey matters; and $\sigma_{WM}$ and $\sigma_{GM}$ are the standard deviations of parameter values in white and grey matters. Higher TC values indicate greater tissue contrast.

## 4 RESULTS

### 4.1 Simulation results based on the tissue model

In Figure 5(a), we present the simulated diffusion signal based on the tissue model presented in Section 3.1 for various axon radii R, showing our in vivo dMRI acquisition setting (Gmax =80 mT/m, $\delta = 21.6\ ms$, and $\Delta = 31.9\ ms$) is sufficient to distinguish radii between $3.5\ \mu m$ and $9.5\ \mu m$ in steps of 1 $\mu m$. Smaller radii between 0.5 $\mu m$ and 2.5 $\mu m$ are more difficult to distinguish from one another. By setting $\delta$ in the simulation to a smaller value, we only observe a marginal improvement in sensitivity to small axon radii. In Figure 5(b), we illustrate the fitting of anomalous diffusion models (7)-(10) to the simulated signal with $R = 6.5\ \mu m$ at $b = 0, 0.5, 1.5, 2.5, 3.5\ ms/\mu m^2$. We see that anomalous diffusion models are able to fit the simulated signal very well across five b-values. Furthermore, we present scatter plots of fitted diffusion parameters from models (6)-(11) versus various radii in Figure 6. We can see that all the diffusion parameters increase as axon radius increases, indicating that all the diffusion parameters in the CTRW modelling framework are sensitive to the changes in axon radius.

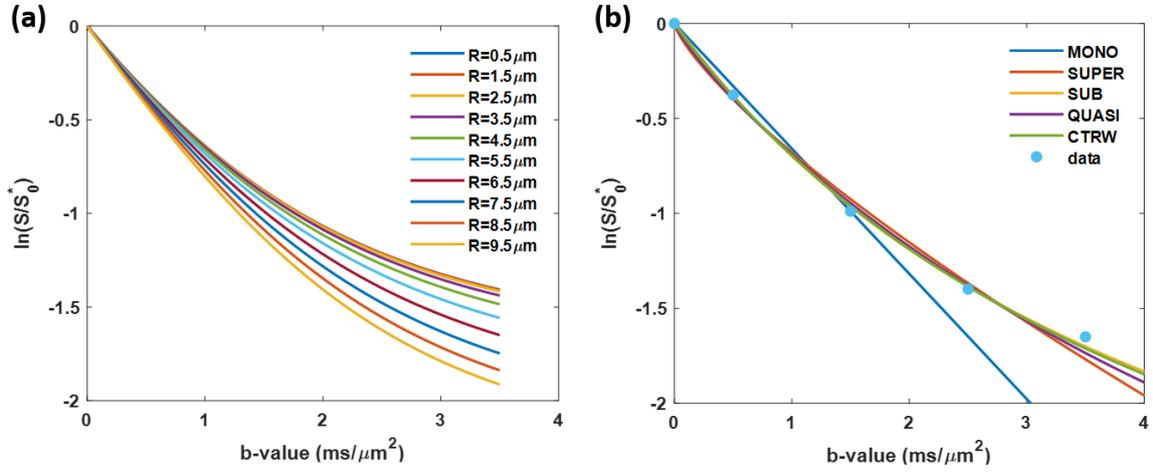

**Figure 5.** Plot of simulated diffusion signal for various mean axon radii R, (a), and fitted signal from models (6)-(10) to the simulated signal (blue dots) with R = 6.5 μm and $b = 0, 0.5, 1.5, 2.5, 3.5\ ms/\mu m^2$, (b).

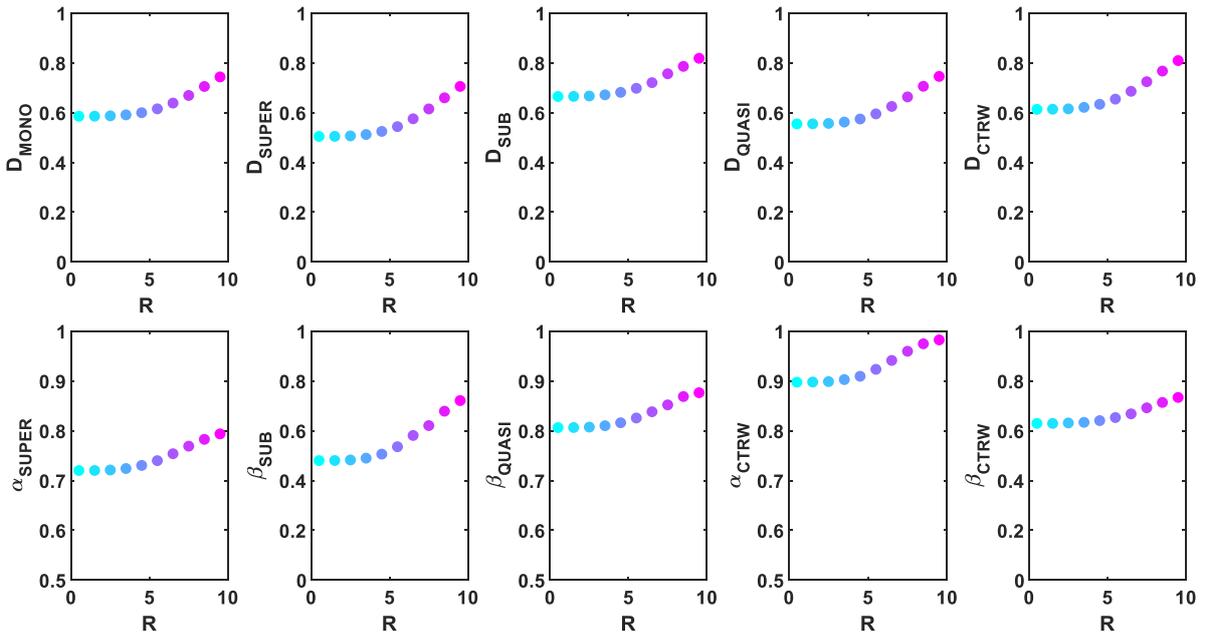

**Figure 6**. Scatter plots of various diffusion parameters (top row: $D_{app}$ for each model; bottom row: $\alpha$ and/or $\beta$ for each model) vs mean axon radius $R$ (varying from 0.5 μm - 9.5 μm in steps of 1 μm). Diffusivities are in units of μm²/ms, $\alpha$ and $\beta$ are in arbitrary units, and mean axon radius $R$ is in units of μm. Points are color-coded by the size of radius.

### 4.2 Maps of anomalous diffusion parameters

In this section, we illustrate the parameter maps and fitting errors that are obtained under the unified CTRW framework. Figure 7 presents the parameter maps for the various diffusion models in Eqs. (6-11) for a healthy male participant of age 41. All the anomalous diffusion parameters provide contrast between white matter, grey matter and CSF. The means and standard deviations of the various diffusion model parameters in the corpus callosum and its subregions across all the subjects are presented in Table 1. The fitting errors were computed as the root-mean-square error (RMSE) for each voxel. Figure 8 presents scatter plots of the RMSE versus fitted parameter values of each model in the corpus callosum of a male subject of age 41. Data points in this figure are colored by density (yellow indicates higher density and blue indicates lower density). No trends were found between the errors and the fitted parameter values, suggesting the errors are random and not correlated with the fitted parameter values.

**Table 1.** Mean ($\bar{\mu}$) and standard deviation ($\sigma$) values computed from various fitted diffusion model parameters in the corpus callosum and its subregions across all the subjects (n=65). Abbreviations: MONO (mono-exponential model), SUPER (super-diffusion model), SUB (sub-diffusion model), QUASI (quasi-diffusion model), CTRW (continuous time random walk model), DKI (diffusional kurtosis imaging).

| | | Rostrum | Genu | Rostral body | Anterior midbody | Posterior midbody | Isthmus | Splenium | Corpus callosum |
|---|---|---|---|---|---|---|---|---|---|
| $D_{MONO}$ (mm²/s) | $\bar{\mu}$ (× 10⁻³) | 0.736 | 0.753 | 0.793 | 0.834 | 0.716 | 0.614 | 0.596 | 0.684 |
| | $\sigma$ (× 10⁻³) | 0.153 | 0.135 | 0.171 | 0.167 | 0.174 | 0.191 | 0.214 | 0.184 |
| $D_{SUPER}$ (mm²/s) | $\bar{\mu}$ (× 10⁻³) | 0.645 | 0.641 | 0.712 | 0.769 | 0.628 | 0.512 | 0.489 | 0.578 |
| | $\sigma$ (× 10⁻³) | 0.195 | 0.164 | 0.249 | 0.261 | 0.244 | 0.207 | 0.253 | 0.218 |
| $D_{SUB}$ (mm²/s) | $\bar{\mu}$ (× 10⁻³) | 0.836 | 0.843 | 0.927 | 1.047 | 0.866 | 0.755 | 0.786 | 0.821 |
| | $\sigma$ (× 10⁻³) | 0.268 | 0.222 | 0.344 | 0.401 | 0.355 | 0.289 | 0.428 | 0.309 |
| $D_{QUASI}$ (mm²/s) | $\bar{\mu}$ (× 10⁻³) | 0.712 | 0.720 | 0.796 | 0.893 | 0.718 | 0.576 | 0.591 | 0.660 |
| | $\sigma$ (× 10⁻³) | 0.237 | 0.208 | 0.296 | 0.332 | 0.315 | 0.261 | 0.403 | 0.286 |
| $D_{CTRW}$ (mm²/s) | $\bar{\mu}$ (× 10⁻³) | 0.841 | 0.899 | 0.969 | 1.114 | 0.891 | 0.723 | 0.748 | 0.831 |
| | $\sigma$ (× 10⁻³) | 0.276 | 0.265 | 0.335 | 0.369 | 0.359 | 0.335 | 0.446 | 0.348 |
| $D^*$ (mm²/s) | $\bar{\mu}$ (× 10⁻³) | 0.925 | 0.941 | 1.028 | 1.155 | 0.956 | 0.836 | 0.855 | 0.908 |
| | $\sigma$ (× 10⁻³) | 0.290 | 0.246 | 0.365 | 0.410 | 0.374 | 0.314 | 0.442 | 0.330 |
| $D_{DKI}$ (mm²/s) | $\bar{\mu}$ (× 10⁻³) | 0.893 | 0.937 | 1.022 | 1.200 | 0.963 | 0.757 | 0.786 | 0.874 |
| | $\sigma$ (× 10⁻³) | 0.354 | 0.324 | 0.403 | 0.435 | 0.447 | 0.417 | 0.526 | 0.422 |
| $\alpha_{SUPER}$ | $\bar{\mu}$ | 0.752 | 0.712 | 0.714 | 0.647 | 0.708 | 0.748 | 0.718 | 0.715 |
| | $\sigma$ | 0.114 | 0.102 | 0.102 | 0.094 | 0.119 | 0.147 | 0.160 | 0.132 |
| $\beta_{SUB}$ | $\bar{\mu}$ | 0.609 | 0.522 | 0.545 | 0.436 | 0.535 | 0.569 | 0.492 | 0.525 |
| | $\sigma$ | 0.182 | 0.125 | 0.127 | 0.159 | 0.201 | 0.245 | 0.293 | 0.229 |
| $\beta_{QUASI}$ | $\bar{\mu}$ | 0.836 | 0.808 | 0.816 | 0.773 | 0.806 | 0.827 | 0.792 | 0.806 |
| | $\sigma$ | 0.069 | 0.061 | 0.059 | 0.057 | 0.075 | 0.097 | 0.126 | 0.087 |
| $\alpha_{CTRW}$ | $\bar{\mu}$ | 0.973 | 0.979 | 0.983 | 0.949 | 0.959 | 0.958 | 0.915 | 0.956 |
| | $\sigma$ | 0.064 | 0.043 | 0.037 | 0.069 | 0.072 | 0.079 | 0.132 | 0.079 |
| $\beta_{CTRW}$ | $\bar{\mu}$ | 0.579 | 0.464 | 0.501 | 0.419 | 0.470 | 0.501 | 0.405 | 0.446 |
| | $\sigma$ | 0.202 | 0.165 | 0.163 | 0.141 | 0.213 | 0.273 | 0.305 | 0.233 |
| $K^*$ | $\bar{\mu}$ | 1.482 | 1.672 | 1.587 | 1.939 | 1.761 | 1.811 | 2.088 | 1.843 |
| | $\sigma$ | 0.480 | 0.397 | 0.416 | 0.506 | 0.492 | 0.523 | 0.584 | 0.520 |
| $K_{DKI}$ | $\bar{\mu}$ | 0.670 | 0.785 | 0.759 | 0.865 | 0.809 | 0.752 | 0.855 | 0.800 |
| | $\sigma$ | 0.380 | 0.336 | 0.317 | 0.228 | 0.368 | 0.503 | 0.577 | 0.426 |

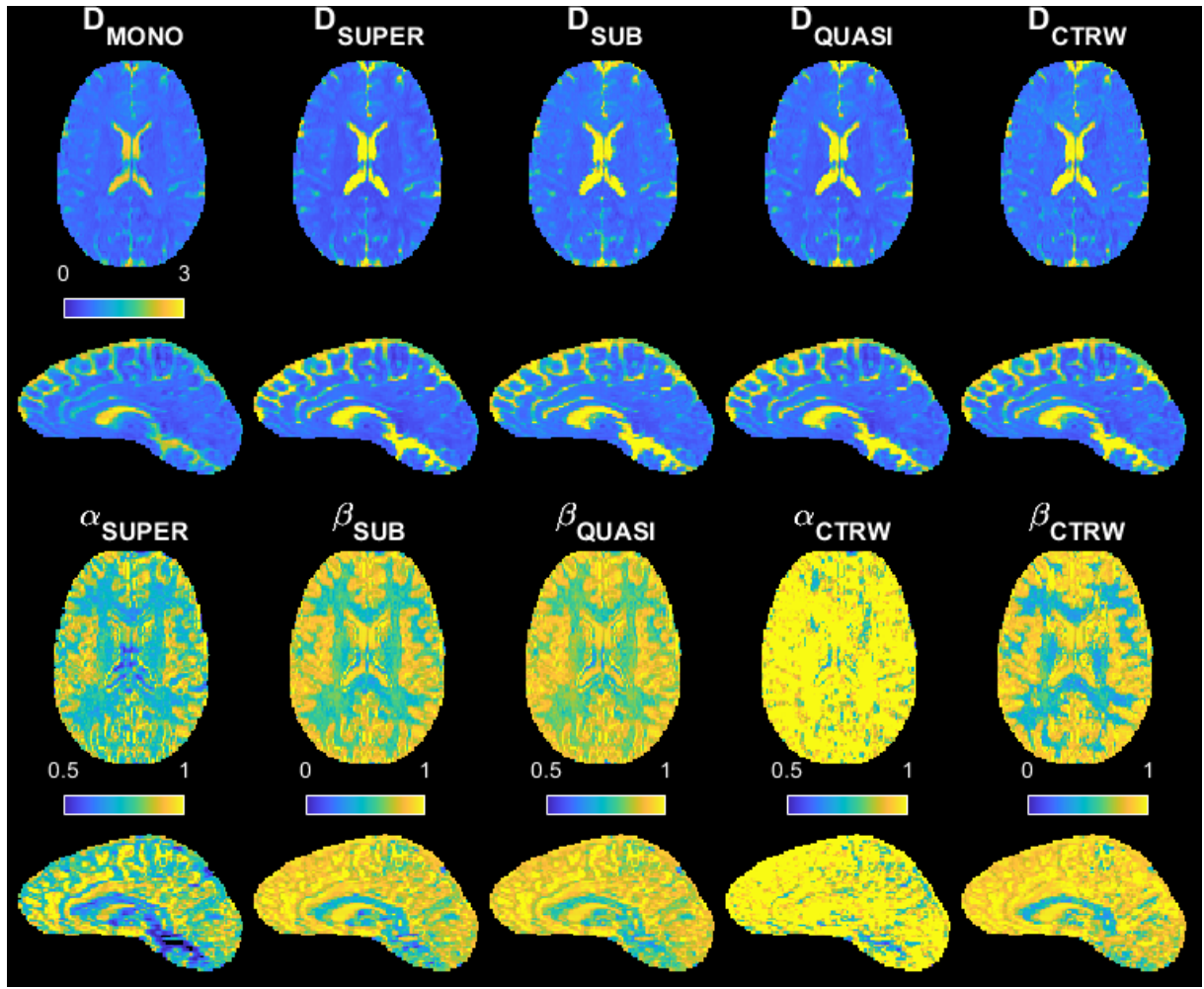

**Figure 7.** An example set of anomalous diffusion parameter maps generated for a healthy male participant of age 41. Parameters $D_{MONO}, D_{SUPER}, D_{SUB}, D_{QUASI}, D_{CTRW}$ (in units of $10^{-3} \times mm^2/s$) are the apparent diffusivities estimated from the models in Eqs.(6)-(11). Parameters $\alpha$ and/or $\beta$ are the spatial and temporal anomalous diffusion metrics for each model. Note, maps for $(D_{FBT}, \alpha_{FBT})$, not shown here, are the same as $(D_{SUPER}, \alpha_{SUPER})$ since they both take the stretched exponential form as explained in section 2.1. Abbreviations: MONO (mono-exponential model), SUPER (super-diffusion model), SUB (sub-diffusion model), QUASI (quasi-diffusion model), CTRW (continuous time random walk model), and FBT (fractional Bloch-Torrey equation).

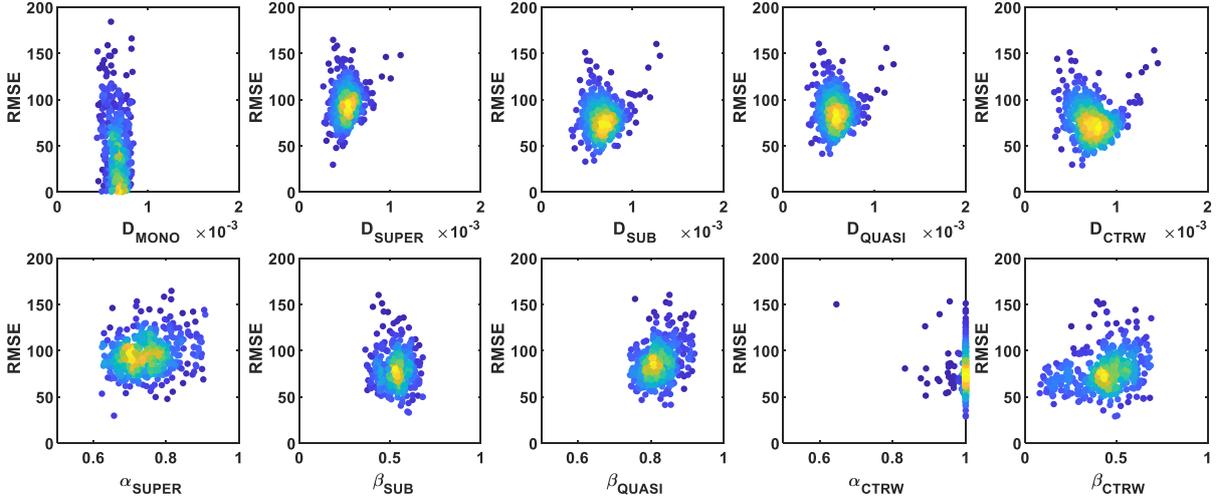

**Figure 8.** Scatter plots of the root-mean-square errors (RMSE) versus fitted parameter values of each model (6)-(11) in the corpus callosum of a male subject of age 41, coloured by the density of the points. Top row: apparent diffusivities from various diffusion models; bottom row: $\alpha$ and $\beta$ from various diffusion models. Abbreviations: MONO (mono-exponential model), SUPER (super-diffusion model), SUB (sub-diffusion model), QUASI (quasi-diffusion model), and CTRW (continuous time random walk model).

### 4.3 Comparison of kurtosis estimated from sub-diffusion and DKI models

Figure 9 shows diffusivity and kurtosis maps computed using two different approaches. In Figure 9(a), the diffusivity $D^*$ and kurtosis $K^*$ were computed based on the sub-diffusion model parameters $D_{SUB}$ and $\beta_{SUB}$ using the full b-value datasets (b = 0, 500, 1500, 2500, 3500 s/mm$^2$). This way of computing kurtosis does not limit the maximum b-value used for the estimation. In Figure 9(b), $D^*$ and $K^*$ were computed over the same b-value range as for the DKI estimates. In Figure 9(c), we provide the diffusivity $D_{DKI}$ and kurtosis $K_{DKI}$ estimated from the standard DKI formulation, Eq. (13), using a nonlinear least square fitting algorithm as in Jensen et al. (2005) with maximum b-value limited to 2500 $s/mm^2$. Grey and white matter contrast in the kurtosis map appears markedly improved for $K^*$ in comparison with $K_{DKI}$, with $TC = 2.83$ for $K^*$ in column (a) and $TC = 0.87$ for $K_{DKI}$ in column (c). Furthermore, in a region marked by red arrows, white matter structure is only visible in the $K^*$ map computed from the full b-value dataset in column (a).

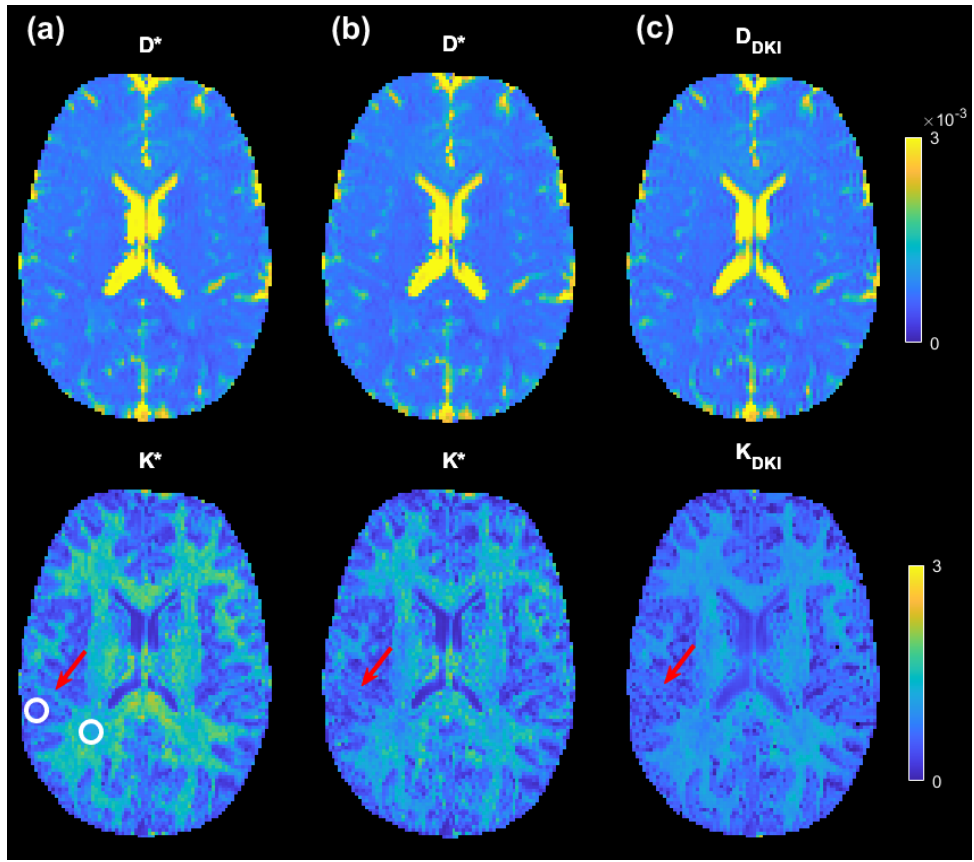

**Figure 9**. Comparison of diffusivity and kurtosis estimated from the sub-diffusion and DKI models for the same subject in Figure 7. Column (a): $D^*$ and $K^*$ computed from sub-diffusion using the full b-value datasets (b = 0, 500, 1500, 2500, 3500 $s/mm^2$); Column (b): $D^*$ and $K^*$ computed using the same b-value range as for the DKI estimates; Column (c): $D_{DKI}$ and $K_{DKI}$ estimated from the DKI model using the datasets with b = 0, 500, 1500, 2500 $s/mm^2$. White circles indicate ROIs for computing tissue contrast between white and grey matters. In a region marked by red arrows, white matter structure is only visible on the $K^*$ maps in column (a) computed from the full b-value datasets.

Figure 10 presents the Bland-Altman plots to analyse the agreement between $K^*$ and $K_{DKI}$ (panel a) and $D^*$ and $D_{DKI}$ (panel b) in the corpus callosum of a male subject of age 41. In Figure 10(a), we see that the mean difference between the two measurements of kurtosis is 0.831, with 95% of differences between 0.182 and 1.480. We also see that there is a significant positive correlation ($r = 0.42, p < 1 \times 10^{-15}$) between the difference and the average of the two measurements, suggesting the difference tends to be bigger as measurement gets bigger. In Figure 10(b), we see that the mean difference between the two measurements of diffusivity is $-3.40 \times 10^{-5}$ mm$^2$/s, with 95% of differences lying between $-2.55 \times 10^{-4}$ mm$^2$/s and $1.87 \times 10^{-4}$ mm$^2$/s. We also see that there is a significant negative correlation ($r = -0.60, p < 1 \times 10^{-15}$) between the difference and the average of the two measurements. The interesting observations here are: (i) when $D^* > D_{DKI}$, the difference tends to be bigger as the measurement gets smaller; and (ii) when $D^* < D_{DKI}$, the difference tends to be bigger as the measurement gets bigger. In summary, the Bland-Altman plots suggest that the diffusivity and kurtosis estimated using the sub-diffusion and DKI models are not the same.

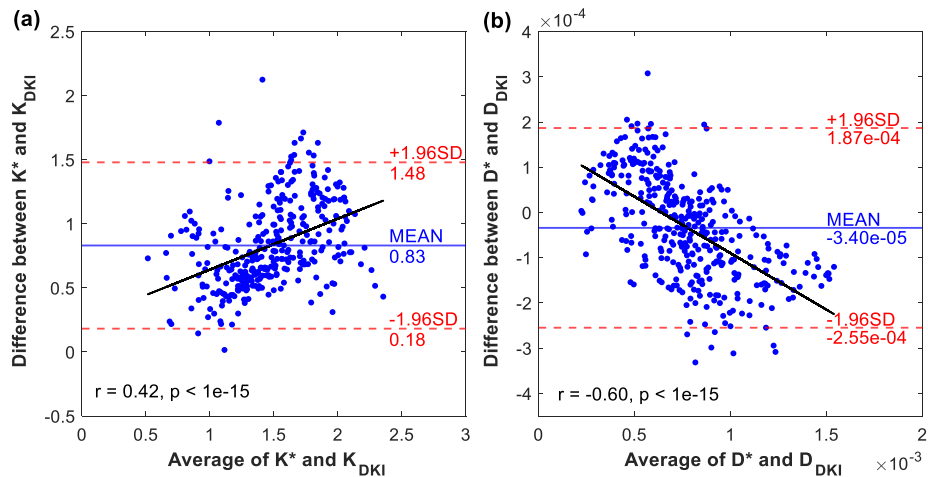

**Figure 10.** Bland-Altman plots to compare diffusivity and kurtosis measured from the sub-diffusion and DKI models in the corpus callosum of a male subject of age 41: (a) the difference between $K^*$ and $K_{DKI}$ versus the average of them; (b) the difference between $D^*$ and $D_{DKI}$ versus the average of them.

### 4.4 Effects of age on anomalous diffusion parameters in corpus callosum

The detailed correlation analysis between age and each diffusion parameter in the corpus callosum and its seven subregions has been provided in Figures S2-S14 in supplementary material. A summary of the linear correlation between age and each diffusion parameter for the corpus callosum and its seven subregions is presented in Table 2. All of the apparent diffusivities showed a significant ($p \leq 0.05/7$) positive correlation with age in the genu. All of the anomalous diffusion indices ($\alpha$ and/or $\beta$) showed a significant ($p \leq 0.05/7$) positive correlation with age in the splenium and whole corpus callosum, except for $\alpha_{CTRW}$. Some $\alpha$ and $\beta$ also indicated significant ($p \leq 0.05/7$) or almost significant ($p \leq 0.05$) relationships with age in the rostral body, anterior midbody and posterior midbody. No significant correlations were found in the rostrum, genu and isthmus.

**Table 2.** Summary of the correlation coefficients (with $p \leq 0.05$) between age and various diffusion model parameters in the corpus callosum and its seven sub-regions. An asterisk (*) indicates the p-value of the corresponding correlation coefficient surviving multiple comparisons correction based on the Bonferroni method (i.e. uncorrected $p \leq 0.05/7$). The correlation coefficient has been colour-coded. Yellow indicates positive correlation, blue indicates negative correlation, and blank indicates correlation is not significant (NS). Abbreviations: MONO (mono-exponential model), SUPER (super-diffusion model), SUB (sub-diffusion model), QUASI (quasi-diffusion model), CTRW (continuous time random walk model), DKI (diffusional kurtosis imaging).

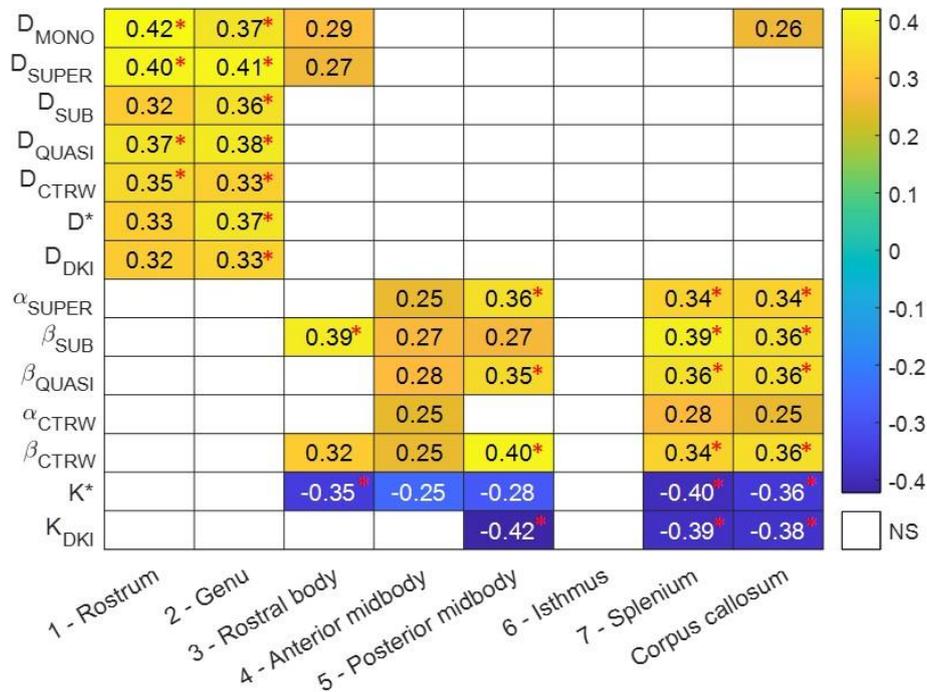

| | 1 - Rostrum | 2 - Genu | 3 - Rostral body | 4 - Anterior midbody | 5 - Posterior midbody | 6 - Isthmus | 7 - Splenium | Corpus callosum |
|---|---|---|---|---|---|---|---|---|
| $D_{MONO}$ | 0.42* | 0.37* | 0.29 | | | | | 0.26 |
| $D_{SUPER}$ | 0.40* | 0.41* | 0.27 | | | | | |
| $D_{SUB}$ | 0.32 | 0.36* | | | | | | |
| $D_{QUASI}$ | 0.37* | 0.38* | | | | | | |
| $D_{CTRW}$ | 0.35* | 0.33* | | | | | | |
| $D^*$ | 0.33 | 0.37* | | | | | | |
| $D_{DKI}$ | 0.32 | 0.33* | | | | | | |
| $\alpha_{SUPER}$ | | | | 0.25 | 0.36* | | 0.34* | 0.34* |
| $\beta_{SUB}$ | | | 0.39* | 0.27 | 0.27 | | 0.39* | 0.36* |
| $\beta_{QUASI}$ | | | | 0.28 | 0.35* | | 0.36* | 0.36* |
| $\alpha_{CTRW}$ | | | | 0.25 | | | 0.28 | 0.25 |
| $\beta_{CTRW}$ | | | 0.32 | 0.25 | 0.40* | | 0.34* | 0.36* |
| $K^*$ | | | -0.35* | -0.25 | -0.28 | | -0.40* | -0.36* |
| $K_{DKI}$ | | | | | -0.42* | | -0.39* | -0.38* |

$K^*$ and $K_{DKI}$ showed a consistent negative correlation with age in the splenium and whole corpus callosum, but different correlations in the midbody sub-regions of corpus callosum. Note here we do not expect the ageing correlation between $K^*$ and $K_{DKI}$ would necessarily be similar, since they are two different approaches to estimate the true kurtosis as shown in section 4.3. The aging correlation of $K^*$ is shown to be similar to $\beta_{SUB}$ (with an opposite sign) but not identical, due to the nonlinear relationship between $K^*$ and $\beta_{SUB}$, Eq.(18), as seen in Figure 2(b).

Quadratic correlation with age was also studied and the $Age^2$ term showed no significant contribution to the regression $Y = a_0 + a_1 * Age + a_2 * Age^2$, where $Y$ represents the fitted diffusion model parameter (see the F-test results in Table S1 in the supplementary material).

## 5 DISCUSSION

The continuous time random walk (CTRW) model and its special cases, a class of non-Gaussian (anomalous) diffusion-weighted imaging techniques, have emerged as advanced methods to evaluate both healthy and diseased tissue microstructure. In this study, we frame the CTRW model and its special cases, including super-diffusion (i.e. stretched exponential), sub-diffusion, quasi-diffusion, and fractional Bloch-Torrey models within a general framework that allows the model parameters ($D_{app}, \alpha, \beta$) for each model to be related and compared. In addition, we provide an alternative way of deriving the DKI model through the Taylor series expansion of the sub-diffusion model. This derivation allows us to compute kurtosis and diffusivity as complementary parameters based on the sub-diffusion model parameters $D_{SUB}$ and $\beta_{SUB}$, without limiting maximum b-value to be 3000 s/mm$^2$. We have shown that this way of computing kurtosis ($K^*$) leads to superior white and grey matter contrast compared to the conventional DKI metric $K_{DKI}$. As an application of this class of anomalous diffusion models, we then evaluated how the various anomalous diffusion model parameters varied across different ages (65 participants, 19-78 years of age) in the human corpus callosum, the primary white matter structure within the brain. Results revealed that the anomalous diffusion indices $\alpha$ and $\beta$ correlated with age in the whole corpus callosum (especially splenium), while the apparent diffusivities $D_{app}$ correlated with age in the rostrum and genu (but not the whole corpus callosum). Our findings indicate that the anomalous diffusion indices $\alpha$ and $\beta$ may contain extra information about white matter microstructural changes during aging compared to the apparent diffusivities.

### 5.1 Age-related changes in anomalous diffusion parameters

Diffusion tensor imaging metrics obtained in the corpus callosum are not expected to be different due to gender (Aboitiz et al., 1992; Aboitiz et al., 1996; Hasan et al., 2005; Ota et al., 2006). In terms of aging, it has been shown that diffusivity increases, whilst fractional anisotropy decreases with age (Barrick et al., 2010; Fenoll et al., 2017; Ota et al., 2006; Pfefferbaum et al., 2000). Given that fractional anisotropy is a measure of how anisotropic diffusion is within white matter fibre bundles, a decrease in this parameter suggests that diffusivity is becoming more isotropic. This can occur due to an increase in axon radius, as the apparent radial diffusivity becomes larger, or can be due to loss of fibre density. An overall increase in diffusivity was in agreement with a decrease in fractional anisotropy, since mean diffusivity is simply the average of the diffusivities in the three principal directions and an increase in any direction leads to an overall increase in diffusivity. Our results also indicated a positive trend with age in the genu in the value of apparent diffusivity derived from various anomalous diffusion models (see Table 2).

White matter structure is known to vary across the corpus callosum sub-regions (Aboitiz et al., 1992), and anomalous diffusion indices have shown sensitivity to variations in axon radii in simulations (see section 4.1) and in in vivo dMRI studies (Yu et al., 2018, 2017). Moreover, anomalous diffusion model parameters have been suggested to provide complimentary information to diffusion tensor imaging metrics, and may contain information on myelin and iron in tissue (Caporale et al., 2017; Guerreri et al., 2019). We suspect that the additional information gained is via the spatial anomalous diffusion parameter ($\alpha$ in our case, and $\gamma$ in Caporale et al. (2017) and Guerreri et al. (2019)) since there appears to be a clear correlation between the variance shared by radial diffusivity (a change affecting fractional anisotropy) and the spatial anomalous diffusion index $\alpha$. Others have also proposed that the temporal exponent (Palombo et al., 2011), i.e. $\beta$ in our case, is influenced by the presence of inhomogeneous magnetic fields introduced by spatial variations in tissue microstructure. Further investigations remain needed on the biophysical interpretation of anomalous diffusion model parameters for future work.

Previous study (Guerreri et al., 2019) found that the $\gamma$-metric (i.e., $\alpha$ in our study) from the stretched exponential model (i.e., super-diffusion model) is positively correlated to aging in genu. To our best knowledge, this is the only study to date on the correlation between aging and anomalous diffusion model parameter. Our study investigated age-related changes in five existing anomalous diffusion models including the one in Guerreri et al. (2019), and generalised them under the continuous time random walk framework with a consistent parameter fitting strategy applied. Our results showed that age is positively correlated with all the anomalous diffusion metrics ($D_{app}, \alpha, \beta$) across the whole corpus callosum and especially in splenium (see Table 2). Although the age ranges were similar between our study and Guerreri et al. (2019), we did not find significant correlation between age and $\alpha_{SUPER}$ (i.e., the $\gamma$-metric) in genu as in Guerreri et al. (2019). We may attribute the different findings to different sample size (65 subjects in our study vs 32 subjects in Guerreri et al. (2019)), different diffusion times used ($\Delta = 107$ ms in Guerreri et al. (2019) vs $\Delta = 31.9$ ms in our study), different b-value ranges (11 b-values ranging from 200 to 5000 s/mm$^2$ in Guerreri et al. (2019) vs 4 b-values ranging from 500 to 3500 s/mm$^2$) and different spatial resolutions (in-plane $1.8 \times 1.8$ mm$^2$, slice thickness 3 mm at 3T in Guerreri et al. (2019) compared to isotropic resolution $1.6 \times 1.6 \times 1.6$ mm$^3$ at 7T in our study).

Both kurtosis, $K^*$ derived from the sub-diffusion model and $K_{DKI}$ from the conventional DKI model, showed a decline with age, in agreement with previous age-related findings in kurtosis (Gong et al., 2014; Guerreri et al., 2019).

## 5.2 Link between the sub-diffusion and DKI models

It has been reported that accurate estimation of diffusivity and kurtosis through fitting the DKI model is challenging (Chuhutin et al., 2017; Jelescu et al., 2020; Kiselev, 2010; Poot et al., 2010; Rosenkrantz et al., 2015; Szczepankiewicz et al., 2013). One major limitation of DKI is that it is developed based on the Taylor expansion of the logarithm of the signal at $b = 0$, so it is only an accurate representation of the signal in a sufficiently small neighbourhood of zero (Kiselev, 2010). Hence, to estimate diffusivity and kurtosis, Jensen and Helpern (2010) suggested at least three different b-values (such as $b = 0, 1000, 2000$ s/mm$^2$) are needed and the maximum b-value is limited to 2000-3000 s/mm$^2$ in the brain. Other studies reported that the optimal maximum b-value was found to be dependent on the tissue types and specific pathologies, which makes the experimental design optimal for a whole brain challenging (Chuhutin et al., 2017). Furthermore, the procedure for fitting kurtosis and diffusivity tensors is not trivial. A variety of fitting procedures are currently in use. We refer readers to the descriptive and comparative studies for more details about the implementation and comparison of methods (Chuhutin et al., 2017; Veraart et al., 2011a, 2011b).

Here, we highlight the advantages of using the sub-diffusion model to estimate kurtosis and diffusivity over DKI. First, using the link between the sub-diffusion and DKI models (derived in section 2.3), the diffusivity, $D^*$, and kurtosis, $K^*$, can be obtained as complementary parameters based on the sub-diffusion model parameters $D_{SUB}$ and $\beta_{SUB}$ without a limitation on the maximum b-value. Second, this new approach of computing kurtosis using the full b-value dataset may provide a more accurate measurement of kurtosis than the DKI model. This can be seen in Figure 9, where some white matter structure is clearly more visible on the $K^*$ map using the full b-value datasets compared to $K^*$ and $K_{DKI}$ computed using datasets with maximum b-value limited to 2500 s/mm$^2$. In general, we found superior gray-white matter contrast in the $K^*$ map ($TC = 2.83$) compared to the $K_{DKI}$ map ($TC = 0.87$), as shown in Figure 9. The difference between the sub-diffusion $D^*$ and $K^*$ and DKI $D_{DKI}$ and $K_{DKI}$ measures were further highlighted in a Bland-Altman graph (see Figure 10). Future work may see the extension of the method to estimating diffusion and kurtosis tensors.

We note that Ingo et al. (2015) also derived the link between sub-diffusion parameter $\beta$ and kurtosis, Eq. (18), based on the definition of kurtosis, however they did not establish a direct relationship between the two models, i.e., the DKI model Eq. (13) and the sub-diffusion model Eq. (8). In this study, we derived the form of the truncation error term in Eq. (15), showing that the DKI model is actually a degree two approximation of the sub-diffusion model, Eq. (17). Based on this, two links between the model parameters are established (illustrated in Figure 2): (i) the link between kurtosis $K^*$ and sub-diffusion parameter $\beta_{SUB}$, Eq.(18); and (ii) the link between the apparent diffusivities of two models, $D_{SUB}$ and $D^*$, as in Eq.(16). In addition, following the derivation in Eq. (15), higher degree methods beyond the DKI can be developed in the future.

Furthermore, we note that the relationship between the DKI and the sub-diffusion models cannot be extended to the super-diffusion (7), quasi-diffusion (9) or the full CTRW models (10). This is because of the singularity of the derivatives at $b = 0$ when applying Taylor series expansion to the models (7), (9) and (10). Ingo et al. (2015) also pointed out that when the space fractional index $2\alpha < 2$, the second moment diverges for the diffusion propagator, and hence kurtosis cannot be computed according to its definition $K = \langle x^4 \rangle / \langle x^2 \rangle^2 - 3$.

## 5.3 Time-dependent Diffusion

Time-dependent diffusivity has been described in the context of MRI and in the presence of a complex tissue environment for water diffusion. Various models and concepts have been proposed which account for time-dependence of diffusion (Burcaw et al., 2015; Fieremans et al., 2016; Lee et al., 2020a, 2020b; Novikov et al., 2014, 2019). Whilst our motivation here has not been to map time dependent diffusion, we should nonetheless point out that the most general anomalous diffusion model, the CTRW framework, can provide an explanation of the time dependent diffusivity. From Eq. (5), the diffusivity approximated by fitting the CTRW model is of the form $D_{app}(\bar{\Delta}) = (D_{\alpha,\beta}\bar{\Delta}^{\beta-\alpha})^{\frac{1}{\alpha}}$, where $\bar{\Delta} = \Delta - \delta/3$ is the effective diffusion time, and $D_{\alpha,\beta}$ is the generalised anomalous diffusion coefficient in tissue as defined in section 2.1. Hence, when the effective diffusion time $\bar{\Delta}$ is changing in the experiment, time dependent diffusivity can be observed. For example, from Table 1, in white matter, we see that $\alpha_{CTRW}$ is around 0.95 and $\beta_{CTRW}$ is around 0.45, if $\bar{\Delta}$ is changing then we expect the diffusivity $D_{app}$ is changing with $\bar{\Delta}^{\beta-\alpha} \approx \bar{\Delta}^{-0.5}$, i.e., $D_{app}$ is decreasing with increasing diffusion time $\bar{\Delta}$, as seen in Fieremans et al. (2016), Lee et al. (2020a).

Our data in this study was generated using a fixed diffusion time imaging protocol. Therefore, we are unable to infer on time dependence of diffusion or how it changes with aging based on anomalous diffusion models evaluated. Nonetheless, it would be an interesting experiment to evaluate the influence of a changing $\bar{\Delta}$ withing the CTRW framework. Moreover, $\alpha = \beta$ leads to the quasi-diffusion model, where the mean squared displacement grows linearly with time as in the mono-exponential model. Notably, in this case the contribution from $\bar{\Delta}$ is removed, and it appears that for this model the way *b*- values are set is not expected to impact diffusion time dependence. This conclusion is only observational based on the mathematical expressions presented and should be evaluated experimentally as future work.

## 6 CONCLUSION

We unify several established anomalous diffusion models (including the super-, sub-, and quasi-diffusion models and fractional Bloch-Torrey equation) under the continuous time random walk modelling framework, which provides measures of both apparent diffusivity ($D_{app}$) and anomalous diffusion index ($\alpha$ in space and $\beta$ in time), a measure of the level of heterogeneity or complexity in tissue microstructure. From this novel unified insight, we assess age-related changes in the apparent diffusivity and anomalous diffusion index in the sub-regions of human corpus callosum for a cohort of 65 healthy participants (aged 19~78 years). We found that the apparent diffusivities from anomalous diffusion models show significant positive correlation with age in the rostrum and genu of corpus callosum. Anomalous diffusion indices consistently show significant positive correlation with age in the splenium, where significant correlation is usually not found in previous studies. This finding suggests that anomalous diffusion indices may provide extra information on white matter microstructural changes with aging.

Additionally, we derived that the DKI model is a degree two approximation of the sub-diffusion model. This link between the DKI and sub-diffusion models provides an explicit and maybe more accurate way of computing kurtosis and apparent diffusivity, with important advantages that the maximum b-value is not limited to 3000 s/mm$^2$. Superior tissue contrast is achieved in kurtosis maps based on the sub-diffusion model. Overall, our results suggest that anomalous diffusion models under the CTRW framework play an important role in mapping tissue microstructure and deriving white matter specific micro-parameters.

**ACKNOWLEDGEMENT**

We thank the anonymous reviewers for providing many constructive comments which have greatly helped us improve the manuscript. Q. Yang thanks the Australian Research Council (ARC) for funding her Discovery Early Career Research Award (DE150101842). V. Vegh and D. Reutens acknowledge the National Health and Medical Research Council (NHMRC) for funding a project grant (APP1104933) on tissue microstructure imaging; they are members of the Australian Research Council Training Centre for Innovation in Biomedical Imaging Technology (IC170100035). Q. Yang and V. Vegh acknowledge the support of the Australian Research Council Discovery Project Award DP190101889. Data for this project was collected using a National Imaging Facility (NIF) 7T human MRI scanner, housed at the Centre for Advanced Imaging, University of Queensland. We thank Surabhi Sood, Aiman Al Najjar and Nicole Atcheson for helping with participant scans.

**REFERENCES**

Aboitiz, F., Rodríguez, E., Olivares, R., Zaidel, E., 1996. Age-related changes in fibre composition of the human corpus callosum: sex differences. Neuroreport 7, 1761–1764. https://doi.org/10.1097/00001756-199607290-00013

Aboitiz, F., Scheibel, A.B., Fisher, R.S., Zaidel, E., 1992. Fiber composition of the human corpus callosum. Brain Res 598, 143–153. https://doi.org/10.1016/0006-8993(92)90178-c

Alexander, D.C., Hubbard, P.L., Hall, M.G., Moore, E.A., Ptito, M., Parker, G.J.M., Dyrby, T.B., 2010. Orientationally invariant indices of axon diameter and density from diffusion MRI. NeuroImage 52, 1374–1389. https://doi.org/10.1016/j.neuroimage.2010.05.043

Anderson, S.W., Barry, B., Soto, J., Ozonoff, A., O'Brien, M., Jara, H., 2014. Characterizing non-gaussian, high b-value diffusion in liver fibrosis: Stretched exponential and diffusional kurtosis modeling. Journal of Magnetic Resonance Imaging 39, 827–834. https://doi.org/10.1002/jmri.24234

Andersson, J.L.R., Sotiropoulos, S.N., 2016. An integrated approach to correction for off-resonance effects and subject movement in diffusion MR imaging. NeuroImage 125, 1063–1078. https://doi.org/10.1016/j.neuroimage.2015.10.019

Ardekani, S., Kumar, A., Bartzokis, G., Sinha, U., 2007. Exploratory voxel-based analysis of diffusion indices and hemispheric asymmetry in normal aging. Magnetic Resonance Imaging 25, 154–167. https://doi.org/10.1016/j.mri.2006.09.045

Assaf, Y., Basser, P.J., 2005. Composite hindered and restricted model of diffusion (CHARMED) MR imaging of the human brain. Neuroimage 27, 48–58. https://doi.org/10.1016/j.neuroimage.2005.03.042

Assaf, Y., Blumenfeld-Katzir, T., Yovel, Y., Basser, P.J., 2008. AxCaliber: a method for measuring axon diameter distribution from diffusion MRI. Magn Reson Med 59, 1347–1354. https://doi.org/10.1002/mrm.21577

Barrick, T.R., Charlton, R.A., Clark, C.A., Markus, H.S., 2010. White matter structural decline in normal ageing: A prospective longitudinal study using tract-based spatial statistics. NeuroImage 51, 565–577. https://doi.org/10.1016/j.neuroimage.2010.02.033

Barrick, T.R., Spilling, C.A., Hall, M.G., Howe, F.A., 2021. The Mathematics of Quasi-Diffusion Magnetic Resonance Imaging. Mathematics 9, 1763. https://doi.org/10.3390/math9151763

Barrick, T.R., Spilling, C.A., Ingo, C., Madigan, J., Isaacs, J.D., Rich, P., Jones, T.L., Magin, R.L., Hall, M.G., Howe, F.A., 2020. Quasi-diffusion magnetic resonance imaging (QDI): A fast, high b-value diffusion imaging technique. NeuroImage 211, 116606. https://doi.org/10.1016/j.neuroimage.2020.116606

Basser, P.J., Mattiello, J., LeBihan, D., 1994. MR diffusion tensor spectroscopy and imaging. Biophysical Journal 66, 259–267. https://doi.org/10.1016/S0006-3495(94)80775-1

Bennett, K.M., Schmainda, K.M., Bennett, R.T., Rowe, D.B., Lu, H., Hyde, J.S., 2003. Characterization of continuously distributed cortical water diffusion rates with a stretched-exponential model. Magn Reson Med 50, 727–734. https://doi.org/10.1002/mrm.10581

Berman, S., West, K.L., Does, M.D., Yeatman, J.D., Mezer, A.A., 2018. Evaluating g-ratio weighted changes in the corpus callosum as a function of age and sex. NeuroImage, Microstructural Imaging 182, 304–313. https://doi.org/10.1016/j.neuroimage.2017.06.076

Brancato, V., Cavaliere, C., Salvatore, M., Monti, S., 2019. Non-Gaussian models of diffusion weighted imaging for detection and characterization of prostate cancer: a systematic review and meta-analysis. Sci Rep 9. https://doi.org/10.1038/s41598-019-53350-8

Bueno-Orovio, A., Teh, I., Schneider, J., Burrage, K., Grau, V., 2016. Anomalous Diffusion in Cardiac Tissue as an Index of Myocardial Microstructure. IEEE Transactions on Medical Imaging 35, 2200–2207. https://doi.org/10.1109/TMI.2016.2548503

Burcaw, L.M., Fieremans, E., Novikov, D.S., 2015. Mesoscopic structure of neuronal tracts from time-dependent diffusion. NeuroImage 114, 18–37. https://doi.org/10.1016/j.neuroimage.2015.03.061

Caporale, A., Palombo, M., Macaluso, E., Guerreri, M., Bozzali, M., Capuani, S., 2017. The γ-parameter of anomalous diffusion quantified in human brain by MRI depends on local magnetic susceptibility differences. NeuroImage 147, 619–631. https://doi.org/10.1016/j.neuroimage.2016.12.051

Capuani, S., Palombo, M., Gabrielli, A., Orlandi, A., Maraviglia, B., Pastore, F.S., 2013. Spatio-temporal anomalous diffusion imaging: results in controlled phantoms and in excised human meningiomas. Magnetic Resonance Imaging 31, 359–365. https://doi.org/10.1016/j.mri.2012.08.012

Chuhutin, A., Hansen, B., Jespersen, S.N., 2017. Precision and accuracy of diffusion kurtosis estimation and the influence of b-value selection. NMR Biomed 30, 10.1002/nbm.3777. https://doi.org/10.1002/nbm.3777

Clark, C.A., Le Bihan, D., 2000. Water diffusion compartmentation and anisotropy at high b values in the human brain. Magn Reson Med 44, 852–859. https://doi.org/10.1002/1522-2594(200012)44:6<852::aid-mrm5>3.0.co;2-a

Cordero-Grande, L., Christiaens, D., Hutter, J., Price, A.N., Hajnal, J.V., 2019. Complex diffusion-weighted image estimation via matrix recovery under general noise models. NeuroImage 200, 391–404. https://doi.org/10.1016/j.neuroimage.2019.06.039

Fan, Q., Tian, Q., Ohringer, N.A., Nummenmaa, A., Witzel, T., Tobyne, S.M., Klawiter, E.C., Mekkaoui, C., Rosen, B.R., Wald, L.L., Salat, D.H., Huang, S.Y., 2019. Age-related alterations in axonal microstructure in the corpus callosum measured by high-gradient diffusion MRI. Neuroimage 191, 325–336. https://doi.org/10.1016/j.neuroimage.2019.02.036

Fenoll, R., Pujol, J., Esteba-Castillo, S., de Sola, S., Ribas-Vidal, N., García-Alba, J., Sánchez-Benavides, G., Martínez-Vilavella, G., Deus, J., Dierssen, M., Novell-Alsina, R., de la Torre, R., 2017. Anomalous White Matter Structure and the Effect of Age in Down Syndrome Patients. Journal of Alzheimer's Disease 57, 61–70. https://doi.org/10.3233/JAD-161112

Fieremans, E., Burcaw, L.M., Lee, H.-H., Lemberskiy, G., Veraart, J., Novikov, D.S., 2016. In vivo observation and biophysical interpretation of time-dependent diffusion in human white matter. NeuroImage 129, 414–427. https://doi.org/10.1016/j.neuroimage.2016.01.018

Gatto, R.G., Ye, A.Q., Colon-Perez, L., Mareci, T.H., Lysakowski, A., Price, S.D., Brady, S.T., Karaman, M., Morfini, G., Magin, R.L., 2019. Detection of axonal degeneration in a mouse model of Huntington's disease: comparison between diffusion tensor imaging and anomalous diffusion metrics. Magn Reson Mater Phy 32, 461–471. https://doi.org/10.1007/s10334-019-00742-6

Gong, N.-J., Wong, C.-S., Chan, C.-C., Leung, L.-M., Chu, Y.-C., 2014. Aging in deep gray matter and white matter revealed by diffusional kurtosis imaging. Neurobiology of Aging 35, 2203–2216. https://doi.org/10.1016/j.neurobiolaging.2014.03.011

González, R.G., Schaefer, P.W., Buonanno, F.S., Schwamm, L.H., Budzik, R.F., Rordorf, G., Wang, B., Sorensen, A.G., Koroshetz, W.J., 1999. Diffusion-weighted MR Imaging: Diagnostic Accuracy in Patients Imaged within 6 Hours of Stroke Symptom Onset. Radiology 210, 155–162. https://doi.org/10.1148/radiology.210.1.r99ja02155

Grinberg, F., Farrher, E., Ciobanu, L., Geffroy, F., Bihan, D.L., Shah, N.J., 2014. Non-Gaussian Diffusion Imaging for Enhanced Contrast of Brain Tissue Affected by Ischemic Stroke. PLOS ONE 9, e89225. https://doi.org/10.1371/journal.pone.0089225

Guerreri, M., Palombo, M., Caporale, A., Fasano, F., Macaluso, E., Bozzali, M., Capuani, S., 2019. Age-related microstructural and physiological changes in normal brain measured by MRI γ-metrics derived from anomalous diffusion signal representation. NeuroImage 188, 654–667. https://doi.org/10.1016/j.neuroimage.2018.12.044

Hall, M.G., Barrick, T.R., 2008. From diffusion-weighted MRI to anomalous diffusion imaging. Magnetic Resonance in Medicine 59, 447–455. https://doi.org/10.1002/mrm.21453

Hasan, K.M., Gupta, R.K., Santos, R.M., Wolinsky, J.S., Narayana, P.A., 2005. Diffusion tensor fractional anisotropy of the normal-appearing seven segments of the corpus callosum in healthy adults and relapsing-remitting multiple sclerosis patients. Journal of Magnetic Resonance Imaging 21, 735–743. https://doi.org/10.1002/jmri.20296

Ingo, C., Magin, R.L., Colon-Perez, L., Triplett, W., Mareci, T.H., 2014. On Random Walks and Entropy in Diffusion-Weighted Magnetic Resonance Imaging Studies of Neural Tissue. Magn Reson Med 71, 617–627. https://doi.org/10.1002/mrm.24706

Ingo, C., Sui, Y., Chen, Y., Parrish, T.B., Webb, A.G., Ronen, I., 2015. Parsimonious continuous time random walk models and kurtosis for diffusion in magnetic resonance of biological tissue. Front. Phys. 3. https://doi.org/10.3389/fphy.2015.00011

Jelescu, I.O., Palombo, M., Bagnato, F., Schilling, K.G., 2020. Challenges for biophysical modeling of microstructure. Journal of Neuroscience Methods 344, 108861. https://doi.org/10.1016/j.jneumeth.2020.108861

Jensen, J.H., Helpern, J.A., 2010. MRI quantification of non-Gaussian water diffusion by kurtosis analysis. NMR in Biomedicine 23, 698–710. https://doi.org/10.1002/nbm.1518

Jensen, J.H., Helpern, J.A., Ramani, A., Lu, H., Kaczynski, K., 2005. Diffusional kurtosis imaging: The quantification of non-gaussian water diffusion by means of magnetic resonance imaging. Magnetic Resonance in Medicine 53, 1432–1440. https://doi.org/10.1002/mrm.20508

Johansen-Berg, H., Behrens, T.E.J., 2014. Diffusion MRI. Elsevier. https://doi.org/10.1016/C2011-0-07047-3

Jones, D.K., 2010. Diffusion MRI: Theory, Methods, and Applications, Diffusion MRI. Oxford University Press.

Jones, D.K., Horsfield, M.A., Simmons, A., 1999. Optimal strategies for measuring diffusion in anisotropic systems by magnetic resonance imaging. Magnetic Resonance in Medicine 42, 515–525. https://doi.org/10.1002/(SICI)1522-2594(199909)42:3<515::AID-MRM14>3.0.CO;2-Q

Jung, W., Lee, Jingu, Shin, H.-G., Nam, Y., Zhang, H., Oh, S.-H., Lee, Jongho, 2018. Whole brain g-ratio mapping using myelin water imaging (MWI) and neurite orientation dispersion and density imaging (NODDI). NeuroImage, Microstructural Imaging 182, 379–388. https://doi.org/10.1016/j.neuroimage.2017.09.053

Karaman, M.M., Sui, Y., Wang, H., Magin, R.L., Li, Y., Zhou, X.J., 2016. Differentiating Low- and High-Grade Pediatric Brain Tumors Using a Continuous-Time Random-Walk Diffusion Model at High b-Values. Magn Reson Med 76, 1149–1157. https://doi.org/10.1002/mrm.26012

Khoo, M.M.Y., Tyler, P.A., Saifuddin, A., Padhani, A.R., 2011. Diffusion-weighted imaging (DWI) in musculoskeletal MRI: a critical review. Skeletal Radiol 40, 665–681. https://doi.org/10.1007/s00256-011-1106-6

Kiselev, V.G., 2010. The Cumulant Expansion: An Overarching Mathematical Framework For Understanding Diffusion NMR, in: Jones, PhD, D.K. (Ed.), Diffusion MRI. Oxford University Press, pp. 152–168. https://doi.org/10.1093/med/9780195369779.003.0010

Klages, R., Radons, G., Sokolov, I.M., 2008. Anomalous Transport: Foundations and Applications. Weinheim, Germany : WILEY-VCH Verlag GmbH & Co. KGaA. https://doi.org/10.1002/9783527622979

Kodiweera, C., Alexander, A.L., Harezlak, J., McAllister, T.W., Wu, Y.-C., 2016. Age effects and sex differences in human brain white matter of young to middle-aged adults: A DTI, NODDI, and q-space study. NeuroImage 128, 180–192. https://doi.org/10.1016/j.neuroimage.2015.12.033

Landman, B.A., Farrell, J.A.D., Jones, C.K., Smith, S.A., Prince, J.L., Mori, S., 2007. Effects of diffusion weighting schemes on the reproducibility of DTI-derived fractional anisotropy, mean diffusivity, and principal eigenvector measurements at 1.5T. NeuroImage 36, 1123–1138. https://doi.org/10.1016/j.neuroimage.2007.02.056

Lee, H.-H., Papaioannou, A., Kim, S.-L., Novikov, D.S., Fieremans, E., 2020a. A time-dependent diffusion MRI signature of axon caliber variations and beading. Commun Biol 3, 1–13. https://doi.org/10.1038/s42003-020-1050-x

Lee, H.-H., Papaioannou, A., Novikov, D.S., Fieremans, E., 2020b. In vivo observation and biophysical interpretation of time-dependent diffusion in human cortical gray matter. NeuroImage 222, 117054. https://doi.org/10.1016/j.neuroimage.2020.117054

Magin, R., Hall, M.G., Karaman, M.M., Vegh, V., 2020. Fractional Calculus Models of Magnetic Resonance Phenomena: Relaxation and Diffusion. CRB 48. https://doi.org/10.1615/CritRevBiomedEng.2020033925

Magin, R.L., Abdullah, O., Baleanu, D., Zhou, X.J., 2008. Anomalous diffusion expressed through fractional order differential operators in the Bloch–Torrey equation. Journal of Magnetic Resonance 190, 255–270. https://doi.org/10.1016/j.jmr.2007.11.007

Magin, R.L., Ingo, C., Triplett, W., Colon-Perez, L., Mareci, T.H., 2014. Classification of Fractional Order Biomarkers for Anomalous Diffusion Using *q*-Space Entropy. CRB 42. https://doi.org/10.1615/CritRevBiomedEng.2014011027

McAuliffe, M.J., Lalonde, F.M., McGarry, D., Gandler, W., Csaky, K., Trus, B.L., 2001. Medical Image Processing, Analysis and Visualization in clinical research, in: Proceedings 14th IEEE Symposium on Computer-Based Medical Systems. CBMS 2001. Presented at the Proceedings 14th IEEE Symposium on Computer-Based Medical Systems. CBMS 2001, pp. 381–386. https://doi.org/10.1109/CBMS.2001.941749

Metzler, R., Klafter, J., 2000. The random walk's guide to anomalous diffusion: a fractional dynamics approach. Physics Reports 339, 1–77. https://doi.org/10.1016/S0370-1573(00)00070-3

Miller, K.L., Hargreaves, B.A., Gold, G.E., Pauly, J.M., 2004. Steady-state diffusion-weighted imaging of in vivo knee cartilage. Magnetic Resonance in Medicine 51, 394–398. https://doi.org/10.1002/mrm.10696

Neuman, C.H., 1974. Spin echo of spins diffusing in a bounded medium. The Journal of Chemical Physics 60, 4508–4511. https://doi.org/10.1063/1.1680931

Novikov, D.S., Fieremans, E., Jespersen, S.N., Kiselev, V.G., 2019. Quantifying brain microstructure with diffusion MRI: Theory and parameter estimation. NMR in Biomedicine 32, e3998. https://doi.org/10.1002/nbm.3998

Novikov, D.S., Jensen, J.H., Helpern, J.A., Fieremans, E., 2014. Revealing mesoscopic structural universality with diffusion. Proceedings of the National Academy of Sciences 111, 5088–5093. https://doi.org/10.1073/pnas.1316944111

Ota, M., Obata, T., Akine, Y., Ito, H., Ikehira, H., Asada, T., Suhara, T., 2006. Age-related degeneration of corpus callosum measured with diffusion tensor imaging. NeuroImage 31, 1445–1452. https://doi.org/10.1016/j.neuroimage.2006.02.008

Palombo, M., Gabrielli, A., De Santis, S., Cametti, C., Ruocco, G., Capuani, S., 2011. Spatio-temporal anomalous diffusion in heterogeneous media by nuclear magnetic resonance. J. Chem. Phys. 135, 034504. https://doi.org/10.1063/1.3610367

Panagiotaki, E., Chan, R.W., Dikaios, N., Ahmed, H.U., O'Callaghan, J., Freeman, A., Atkinson, D., Punwani, S., Hawkes, D.J., Alexander, D.C., 2015. Microstructural Characterization of Normal and Malignant Human Prostate Tissue With Vascular, Extracellular, and Restricted Diffusion for Cytometry in Tumours Magnetic Resonance Imaging. Investigative Radiology 50, 218–227. https://doi.org/10.1097/RLI.0000000000000115

Pfefferbaum, A., Sullivan, E.V., Hedehus, M., Lim, K.O., Adalsteinsson, E., Moseley, M., 2000. Age-related decline in brain white matter anisotropy measured with spatially corrected echo-planar diffusion tensor imaging. Magnetic Resonance in Medicine 44, 259–268. https://doi.org/10.1002/1522-2594(200008)44:2<259::AID-MRM13>3.0.CO;2-6

Pietrasik, W., Cribben, I., Olsen, F., Huang, Y., Malykhin, N.V., 2020. Diffusion tensor imaging of the corpus callosum in healthy aging: Investigating higher order polynomial regression modelling. NeuroImage 213, 116675. https://doi.org/10.1016/j.neuroimage.2020.116675

Poot, D.H.J., den Dekker, A.J., Achten, E., Verhoye, M., Sijbers, J., 2010. Optimal Experimental Design for Diffusion Kurtosis Imaging. IEEE Transactions on Medical Imaging 29, 819–829. https://doi.org/10.1109/TMI.2009.2037915

Rosenkrantz, A.B., Padhani, A.R., Chenevert, T.L., Koh, D.-M., Keyzer, F.D., Taouli, B., Bihan, D.L., 2015. Body diffusion kurtosis imaging: Basic principles, applications, and considerations for clinical practice. Journal of Magnetic Resonance Imaging 42, 1190–1202. https://doi.org/10.1002/jmri.24985

Salat, D.H., Tuch, D.S., Greve, D.N., van der Kouwe, A.J.W., Hevelone, N.D., Zaleta, A.K., Rosen, B.R., Fischl, B., Corkin, S., Rosas, H.D., Dale, A.M., 2005. Age-related alterations in white matter microstructure measured by diffusion tensor imaging. Neurobiology of Aging 26, 1215–1227. https://doi.org/10.1016/j.neurobiolaging.2004.09.017

Smith, S.M., Jenkinson, M., Woolrich, M.W., Beckmann, C.F., Behrens, T.E.J., Johansen-Berg, H., Bannister, P.R., De Luca, M., Drobnjak, I., Flitney, D.E., Niazy, R.K., Saunders, J., Vickers, J., Zhang, Y., De Stefano, N., Brady, J.M., Matthews, P.M., 2004. Advances in functional and structural MR image analysis and implementation as FSL. NeuroImage, Mathematics in Brain Imaging 23, S208–S219. https://doi.org/10.1016/j.neuroimage.2004.07.051

Stahon, K.E., Bastian, C., Griffith, S., Kidd, G.J., Brunet, S., Baltan, S., 2016. Age-Related Changes in Axonal and Mitochondrial Ultrastructure and Function in White Matter. J. Neurosci. 36, 9990–10001. https://doi.org/10.1523/JNEUROSCI.1316-16.2016

Sui, Y., Wang, H., Liu, G., Damen, F.W., Wanamaker, C., Li, Y., Zhou, X.J., 2015. Differentiation of Low- and High-Grade Pediatric Brain Tumors with High b-Value Diffusion-weighted MR Imaging and a Fractional Order Calculus Model. Radiology 277, 489–496. https://doi.org/10.1148/radiol.2015142156

Szafer, A., Zhong, J., Gore, J.C., 1995. Theoretical Model for Water Diffusion in Tissues. Magnetic Resonance in Medicine 33, 697–712. https://doi.org/10.1002/mrm.1910330516

Szczepankiewicz, F., Lätt, J., Wirestam, R., Leemans, A., Sundgren, P., van Westen, D., Ståhlberg, F., Nilsson, M., 2013. Variability in diffusion kurtosis imaging: Impact on study design, statistical power and interpretation. NeuroImage 76, 145–154. https://doi.org/10.1016/j.neuroimage.2013.02.078

Tang, L., Zhou, X.J., 2019. Diffusion MRI of cancer: From low to high b-values. Journal of Magnetic Resonance Imaging 49, 23–40. https://doi.org/10.1002/jmri.26293

Tang, Y., Nyengaard, J.R., Pakkenberg, B., Gundersen, H.J.G., 1997. Age-Induced White Matter Changes in the Human Brain: A Stereological Investigation. Neurobiology of Aging 18, 609–615. https://doi.org/10.1016/S0197-4580(97)00155-3

Taouli, B., Koh, D.-M., 2009. Diffusion-weighted MR Imaging of the Liver. Radiology 254, 47–66. https://doi.org/10.1148/radiol.09090021

Thapaliya, K., Vegh, V., Bollmann, S., Barth, M., 2018. Assessment of microstructural signal compartments across the corpus callosum using multi-echo gradient recalled echo at 7 T. NeuroImage, Microstructural Imaging 182, 407–416. https://doi.org/10.1016/j.neuroimage.2017.11.029

Tournier, J.-D., Smith, R., Raffelt, D., Tabbara, R., Dhollander, T., Pietsch, M., Christiaens, D., Jeurissen, B., Yeh, C.-H., Connelly, A., 2019. MRtrix3: A fast, flexible and open software framework for medical image processing and visualisation. NeuroImage 202, 116137. https://doi.org/10.1016/j.neuroimage.2019.116137

Van Gelderen, P., Despres, D., Vanzijl, P.C.M., Moonen, C.T.W., 1994. Evaluation of Restricted Diffusion in Cylinders. Phosphocreatine in Rabbit Leg Muscle. Journal of Magnetic Resonance, Series B 103, 255–260. https://doi.org/10.1006/jmrb.1994.1038

Veraart, J., Hecke, W.V., Sijbers, J., 2011a. Constrained maximum likelihood estimation of the diffusion kurtosis tensor using a Rician noise model. Magnetic Resonance in Medicine 66, 678–686. https://doi.org/10.1002/mrm.22835

Veraart, J., Poot, D.H.J., Hecke, W.V., Blockx, I., Linden, A.V. der, Verhoye, M., Sijbers, J., 2011b. More accurate estimation of diffusion tensor parameters using diffusion kurtosis imaging. Magnetic Resonance in Medicine 65, 138–145. https://doi.org/10.1002/mrm.22603

Warach, S., Gaa, J., Siewert, B., Wielopolski, P., Edelman, R.R., 1995. Acute human stroke studied by whole brain echo planar diffusion-weighted magnetic resonance imaging. Annals of Neurology 37, 231–241. https://doi.org/10.1002/ana.410370214

Witelson, S.F., 1989. Hand and sex differences in the isthmus and genu of the human corpus callosum. A postmortem morphological study. Brain 112 ( Pt 3), 799–835. https://doi.org/10.1093/brain/112.3.799

Yang, Q., Puttick, S., Bruce, Z.C., Day, B.W., Vegh, V., 2020. Investigation of Changes in Anomalous Diffusion Parameters in a Mouse Model of Brain Tumour, in: Bonet-Carne, E., Hutter, J., Palombo, M., Pizzolato, M., Sepehrband, F., Zhang, F. (Eds.), Computational Diffusion MRI, Mathematics and Visualization. Springer International Publishing, Cham, pp. 161–172. https://doi.org/10.1007/978-3-030-52893-5_14

Yu, Q., Reutens, D., O'Brien, K., Vegh, V., 2017. Tissue microstructure features derived from anomalous diffusion measurements in magnetic resonance imaging. Human Brain Mapping 38, 1068–1081. https://doi.org/10.1002/hbm.23441

Yu, Q., Reutens, D., Vegh, V., 2018. Can anomalous diffusion models in magnetic resonance imaging be used to characterise white matter tissue microstructure? NeuroImage 175, 122–137. https://doi.org/10.1016/j.neuroimage.2018.03.052

Zhang, H., Schneider, T., Wheeler-Kingshott, C.A., Alexander, D.C., 2012. NODDI: practical in vivo neurite orientation dispersion and density imaging of the human brain. Neuroimage 61, 1000–1016. https://doi.org/10.1016/j.neuroimage.2012.03.072

Zhong, Z., Merkitch, D., Karaman, M.M., Zhang, J., Sui, Y., Goldman, J.G., Zhou, X.J., 2019. High-Spatial-Resolution Diffusion MRI in Parkinson Disease: Lateral Asymmetry of the Substantia Nigra. Radiology 291, 149–157. https://doi.org/10.1148/radiol.2019181042

# Supplementary Material

**Table S1**. The p-values of the F-test for determining the significance of the linear term ($Age$) and the quadratic term ($Age^2$) in the regression $Y = a_0 + a_1 * Age + a_2 * Age^2$, where $Y$ represents the fitted diffusion model parameter. The p-values (p<0.05) are denoted with an asterisk (*).

|  |  | Rostrum | Genu | Rostral body | Anterior midbody | Posterior midbody | Isthmus | Splenium | Corpus callosum |
|---|---|---|---|---|---|---|---|---|---|
| $D_{MONO}$ | $Age$ | 0.0005* | 0.0025* | 0.0201* | 0.8482 | 0.3723 | 0.0710 | 0.3803 | 0.0378* |
|  | $Age^2$ | 0.2841 | 0.5690 | 0.4500 | 0.9339 | 0.5794 | 0.6515 | 0.7388 | 0.9636 |
| $D_{SUPER}$ | $Age$ | 0.0007* | 0.0008* | 0.0337* | 0.6870 | 0.5030 | 0.2344 | 0.3189 | 0.0792 |
|  | $Age^2$ | 0.1101 | 0.2851 | 0.5831 | 0.9724 | 0.5750 | 0.3510 | 0.5455 | 0.8668 |
| $D_{SUB}$ | $Age$ | 0.0078* | 0.0026* | 0.1448 | 0.4513 | 0.5057 | 0.6636 | 0.7859 | 0.4638 |
|  | $Age^2$ | 0.0699 | 0.1435 | 0.6025 | 0.9994 | 0.5470 | 0.1714 | 0.4957 | 0.8019 |
| $D_{QUASI}$ | $Age$ | 0.0023* | 0.0016* | 0.0520 | 0.4518 | 0.3132 | 0.4153 | 0.6027 | 0.2671 |
|  | $Age^2$ | 0.1242 | 0.2951 | 0.5714 | 0.9828 | 0.5284 | 0.3275 | 0.5009 | 0.8936 |
| $D_{CTRW}$ | $Age$ | 0.0044* | 0.0076* | 0.1281 | 0.3299 | 0.0875 | 0.4935 | 0.8866 | 0.7245 |
|  | $Age^2$ | 0.1299 | 0.8803 | 0.2998 | 0.9109 | 0.4167 | 0.5942 | 0.4239 | 0.4075 |
| $D^*$ | $Age$ | 0.0069* | 0.0022* | 0.1217 | 0.4874 | 0.5046 | 0.5879 | 0.6616 | 0.3576 |
|  | $Age^2$ | 0.0749 | 0.2196 | 0.5641 | 0.9874 | 0.5371 | 0.2020 | 0.5037 | 0.8516 |
| $D_{DKI}$ | $Age$ | 0.0091* | 0.0069* | 0.1909 | 0.2865 | 0.0624 | 0.4566 | 0.6581 | 0.9972 |
|  | $Age^2$ | 0.1419 | 0.9633 | 0.2050 | 0.9321 | 0.4504 | 0.5196 | 0.5117 | 0.5454 |
| $\alpha_{SUPER}$ | $Age$ | 0.5518 | 0.5937 | 0.4079 | 0.0451* | 0.0031* | 0.8148 | 0.0063* | 0.0046* |
|  | $Age^2$ | 0.6727 | 0.0555 | 0.0337* | 0.5131 | 0.3121 | 0.4634 | 0.3719 | 0.0336 |
| $\beta_{SUB}$ | $Age$ | 0.0827 | 0.0506 | 0.0013* | 0.0282* | 0.0303* | 0.0649 | 0.0013* | 0.0031* |
|  | $Age^2$ | 0.4889 | 0.0927 | 0.1304 | 0.7155 | 0.5613 | 0.4090 | 0.9531 | 0.2280 |
| $\beta_{QUASI}$ | $Age$ | 0.9027 | 0.7973 | 0.1055 | 0.0230* | 0.0046* | 0.6508 | 0.0031* | 0.0028* |
|  | $Age^2$ | 0.9875 | 0.0459 | 0.0435* | 0.4771 | 0.3779 | 0.3501 | 0.6867 | 0.0524 |
| $\alpha_{CTRW}$ | $Age$ | 0.9721 | 0.9600 | 0.4015 | 0.0493* | 0.1693 | 0.2915 | 0.0244* | 0.0451* |
|  | $Age^2$ | 0.1409 | 0.7271 | 0.9263 | 0.8860 | 0.7858 | 0.6354 | 0.7178 | 0.9498 |
| $\beta_{CTRW}$ | $Age$ | 0.4092 | 0.0775 | 0.0079* | 0.0445* | 0.0009* | 0.2793 | 0.0060* | 0.0027* |
|  | $Age^2$ | 0.5237 | 0.0041* | 0.0697 | 0.2966 | 0.3429 | 0.0599 | 0.3524 | 0.0194* |
| $K^*$ | $Age$ | 0.2086 | 0.0917 | 0.0037* | 0.0459* | 0.0232* | 0.0803 | 0.0012* | 0.0032* |
|  | $Age^2$ | 0.4930 | 0.2443 | 0.1761 | 0.8217 | 0.3980 | 0.5858 | 0.6004 | 0.2777 |
| $K_{DKI}$ | $Age$ | 0.1584 | 0.6748 | 0.0499* | 0.0582 | 0.0004* | 0.2173 | 0.0016* | 0.0015* |
|  | $Age^2$ | 0.2673 | 0.0102* | 0.0529 | 0.1067 | 0.1605 | 0.1054 | 0.6177 | 0.0457* |

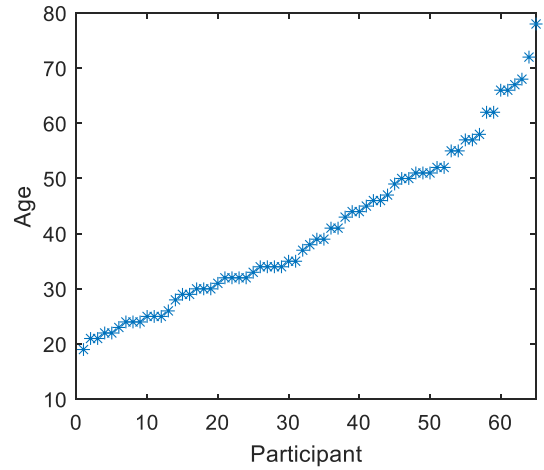

**Figure S1.** The age distribution of the 65 participants.

**Results on the correlation between anomalous diffusion parameters and age**

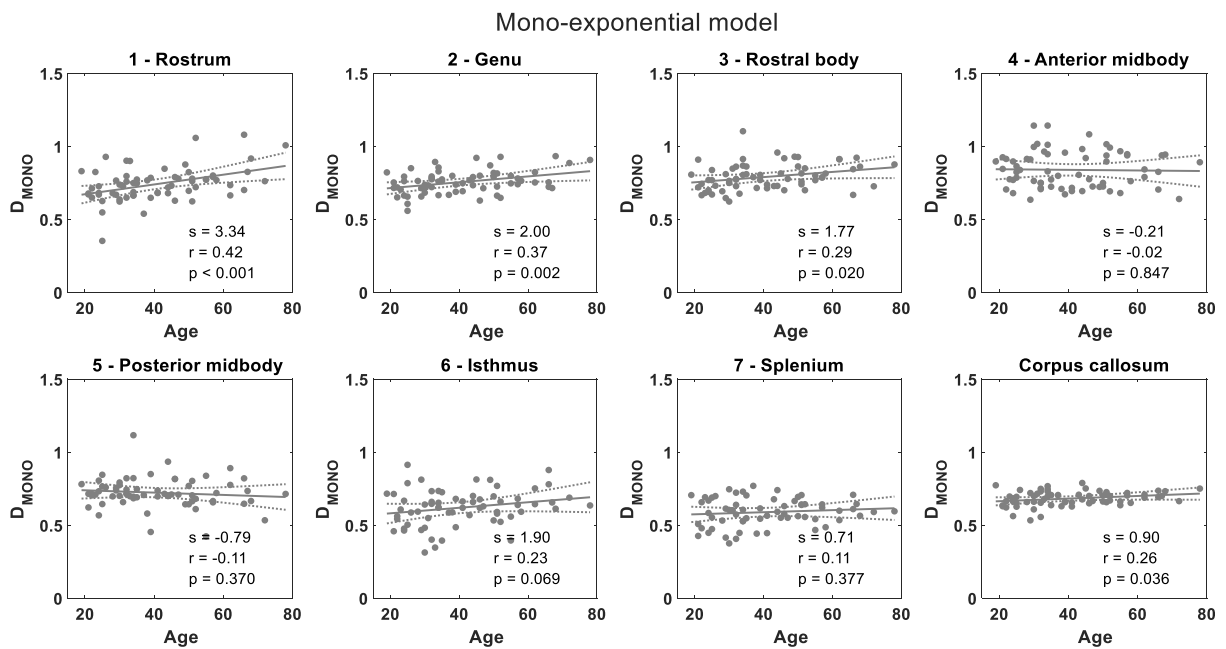

**Figure S2**. Correlation between age and $D_{MONO}$ (in units of $10^{-3} \cdot mm^2/s$) from the mono-exponential model within the corpus callosum and its seven subregions. Dots are the mean $D_{MONO}$ in each ROI for each participant. Solid lines represent the linear trend between $D_{MONO}$ and age. Dashed lines represent the 95% confidence interval. The slope of regression (s in units of $10^{-6} mm^2/s \cdot year^{-1}$), correlation coefficients (r) and the significance level of the linear relationship (p-values) are reported.

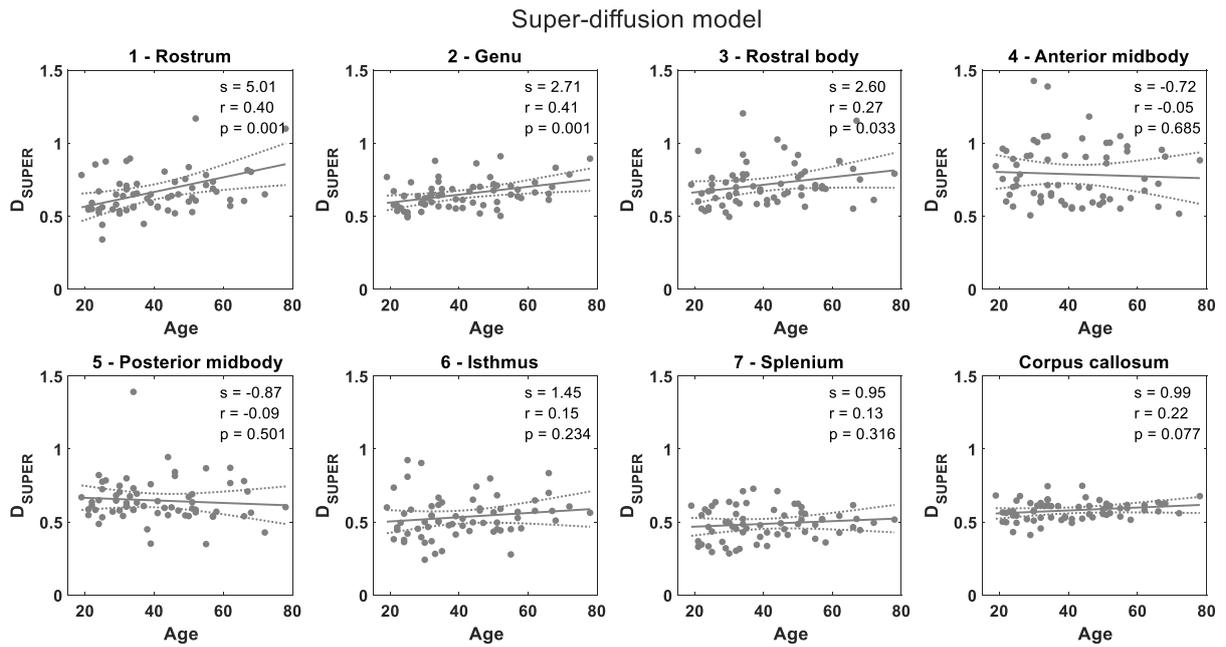

**Figure S3**. Association between age and $D_{SUPER}$ (in units of $10^{-3} \cdot mm^2/s$) from the sub-diffusion model within the corpus callosum and its seven subregions. Dots are the mean $D_{SUPER}$ in each ROI for each participant. Solid lines represent the linear trend between $D_{SUPER}$ and age. Dashed lines represent the 95% confidence interval. The slope of regression (s in units of $10^{-3} \cdot year^{-1}$), correlation coefficients (r) and the significance level of the linear relationship (p-values) are reported.

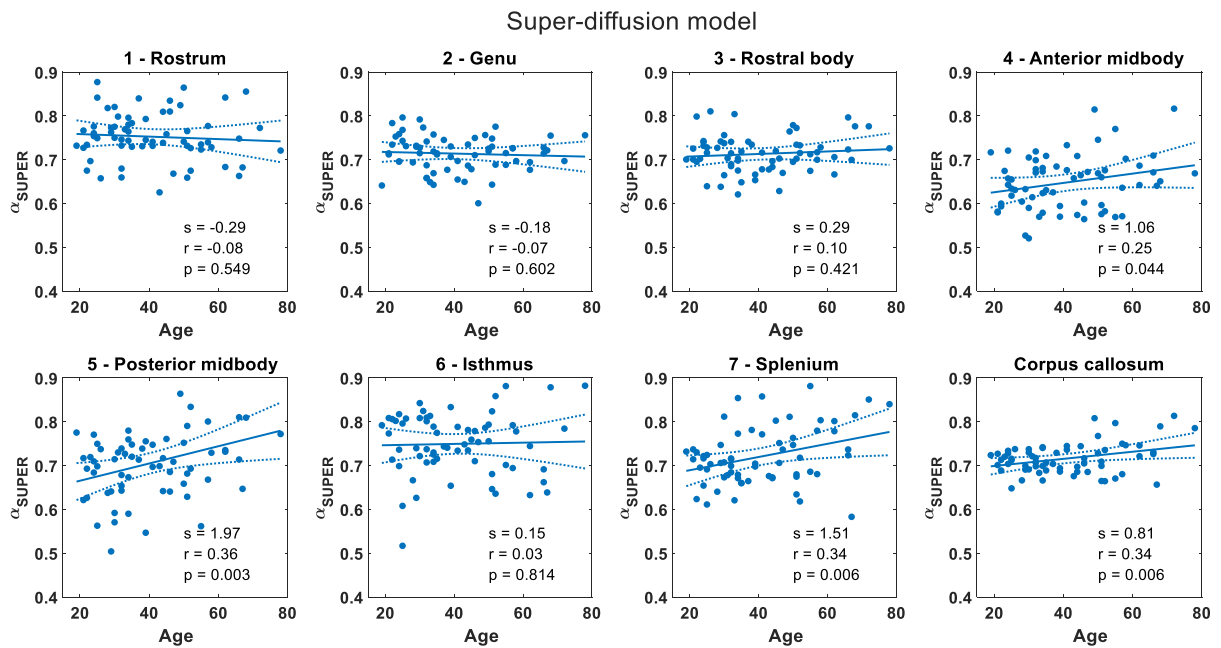

**Figure S4**. Association between age and $\alpha_{SUPER}$ from the super-diffusion model within the corpus callosum and its seven subregions. Dots are the mean $\alpha_{SUPER}$ in each ROI for each participant. Solid lines represent the linear trend between $\alpha_{SUPER}$ and age. Dashed lines represent the 95% confidence interval. The slope of regression (s in units of $10^{-3} \cdot year^{-1}$), correlation coefficients (r) and the significance level of the linear relationship (p-values) are reported.

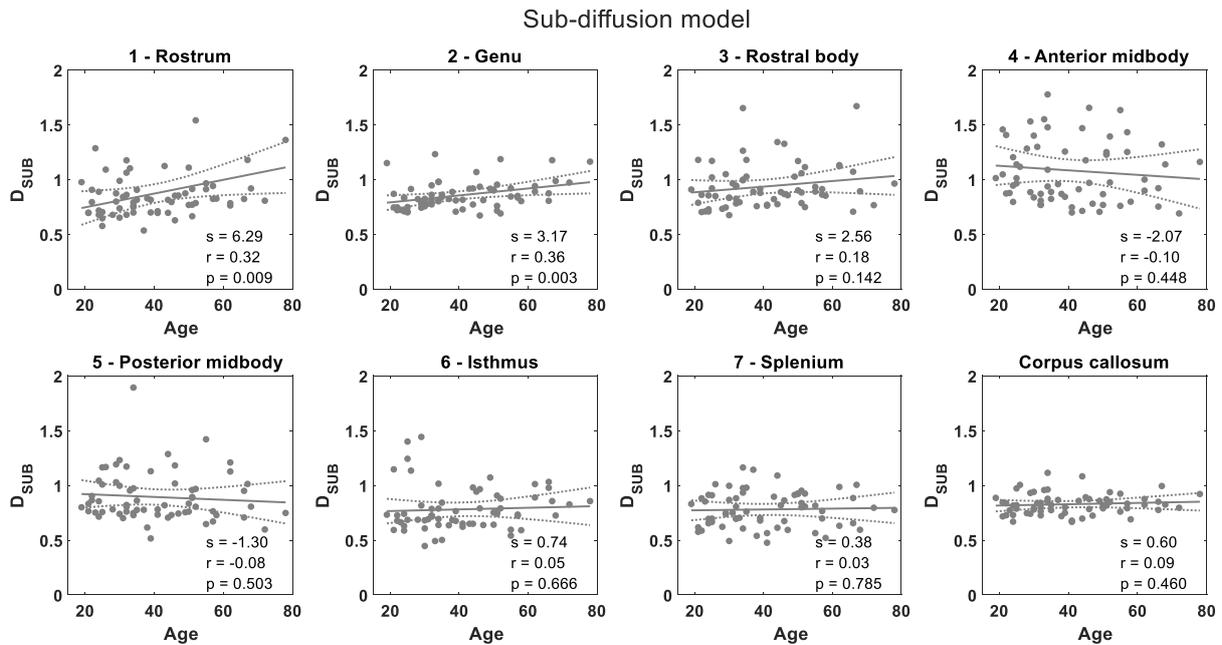

**Figure S5**. Association between age and $D_{SUB}$ (in units of $10^{-3} \cdot mm^2/s$) from the sub-diffusion model within the corpus callosum and its seven subregions. Dots are the mean $D_{SUB}$ in each ROI for each participant. Solid lines represent the linear trend between $D_{SUB}$ and age. Dashed lines represent the 95% confidence interval. The slope of regression (s in units of $10^{-3} \cdot year^{-1}$), correlation coefficients (r) and the significance level of the linear relationship (p-values) are reported.

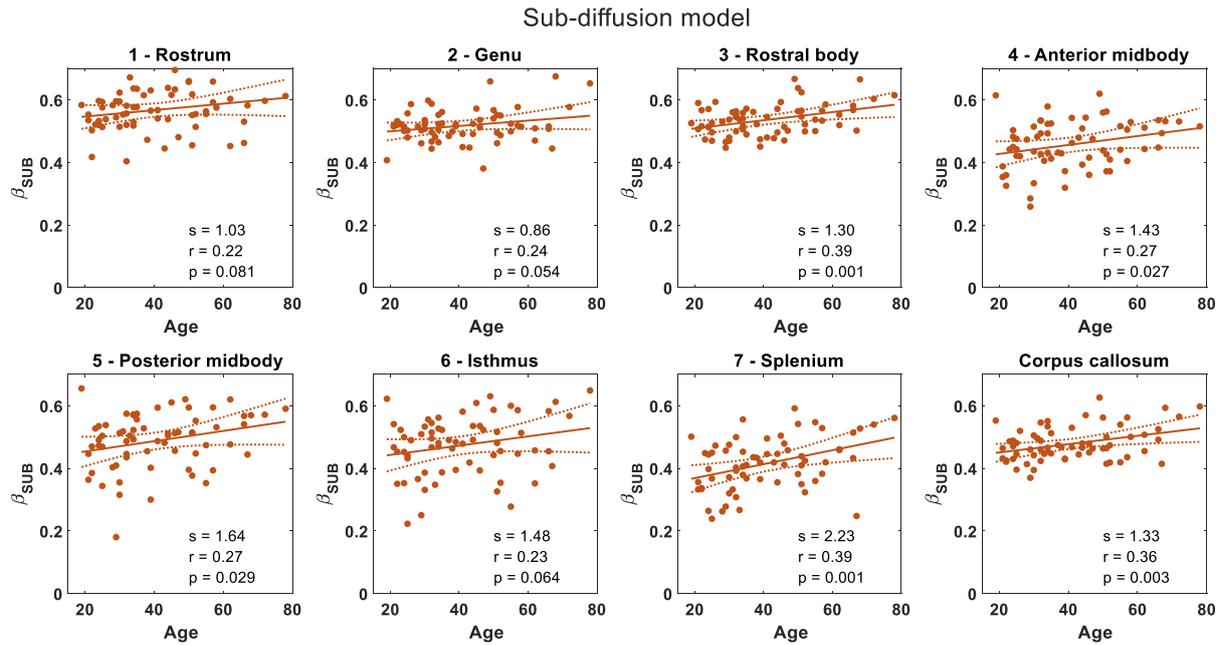

**Figure S6**. Correlation between age and $\beta_{SUB}$ from the sub-diffusion model within the corpus callosum and its seven subregions. Dots are the mean $\beta_{SUB}$ in each ROI for each participant. Solid lines represent the linear trend between $\beta_{SUB}$ and age. Dashed lines represent the 95% confidence interval. The slope of the regression (s in units of $10^{-3} \cdot year^{-1}$), correlation coefficients (r) and the significance level of the linear relationship (p-values) are reported.

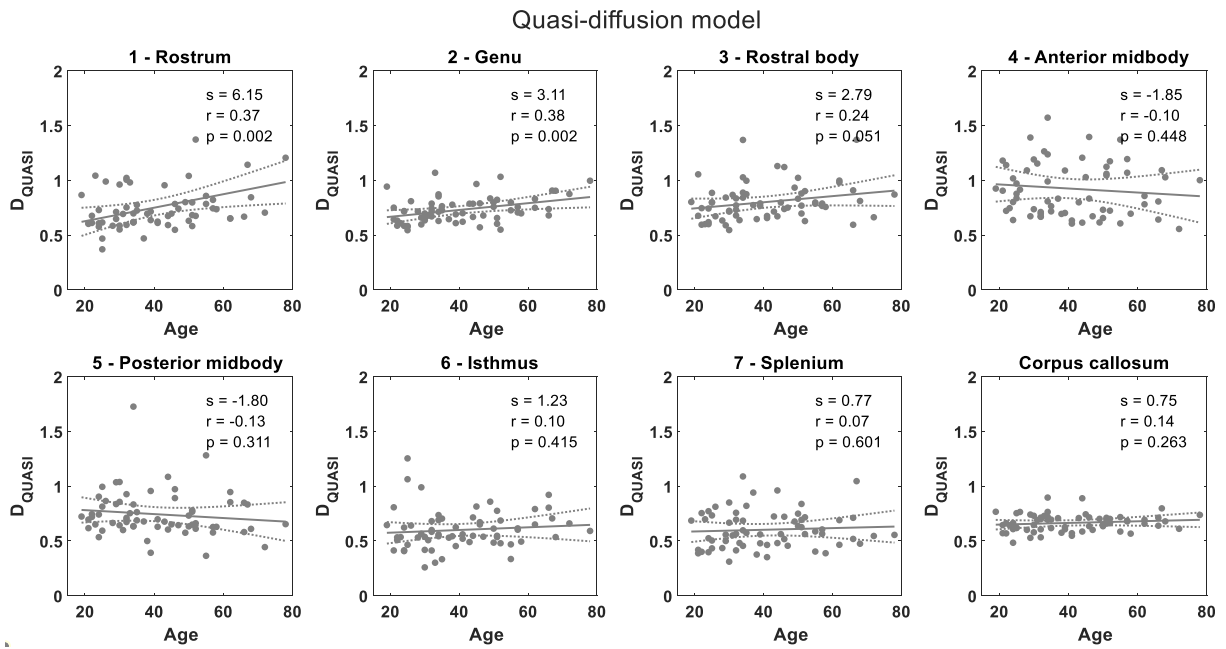

**Figure S7**. Association between age and $D_{QUASI}$ (in units of $10^{-3} \cdot mm^2/s$) from the sub-diffusion model within the corpus callosum and its seven subregions. Dots are the mean $D_{QUASI}$ in each ROI for each participant. Solid

lines represent the linear trend between $D_{QUASI}$ and age. Dashed lines represent the 95% confidence interval. The slope of regression (s in units of $10^{-3} \cdot year^{-1}$), correlation coefficients (r) and the significance level of the linear relationship (p-values) are reported.

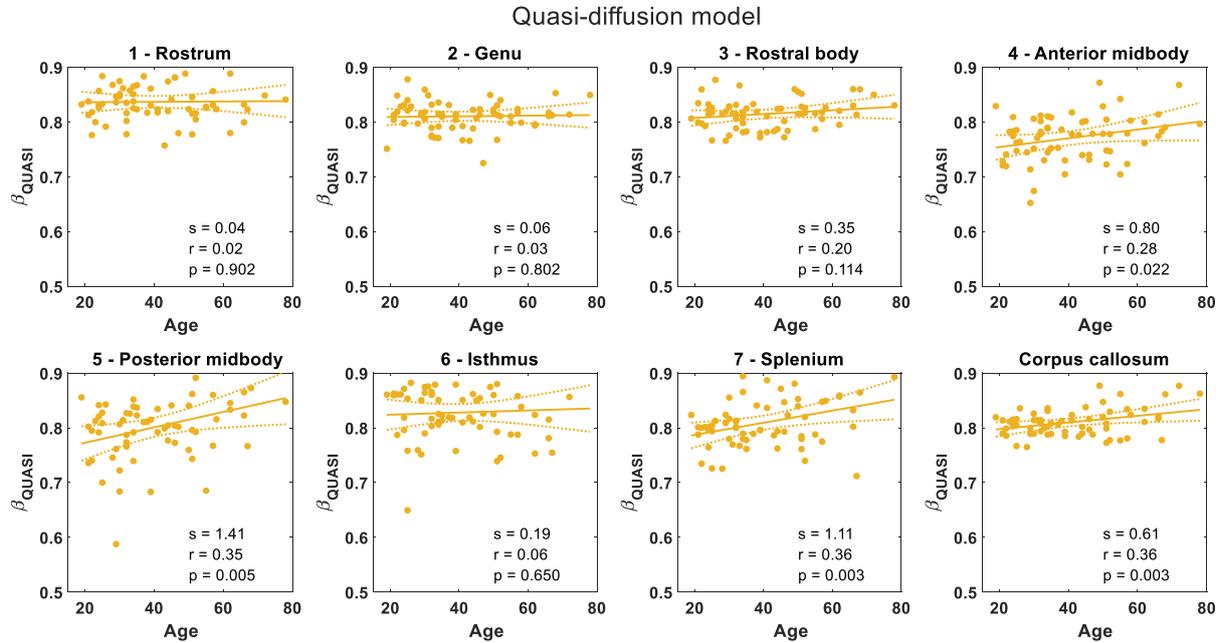

**Figure S8**. Correlation between age and $\beta_{QUASI}$ from the quasi-diffusion model within the corpus callosum and its seven subregions. Dots are the mean $\beta_{QUASI}$ in each ROI for each participant. Solid lines represent the linear trend between $\beta_{QUASI}$ and age. Dashed lines represent the 95% confidence interval. The slope of regression (s in units of $10^{-3} \cdot year^{-1}$), correlation coefficients (r) and the significance level of the linear relationship (p-values) are reported.

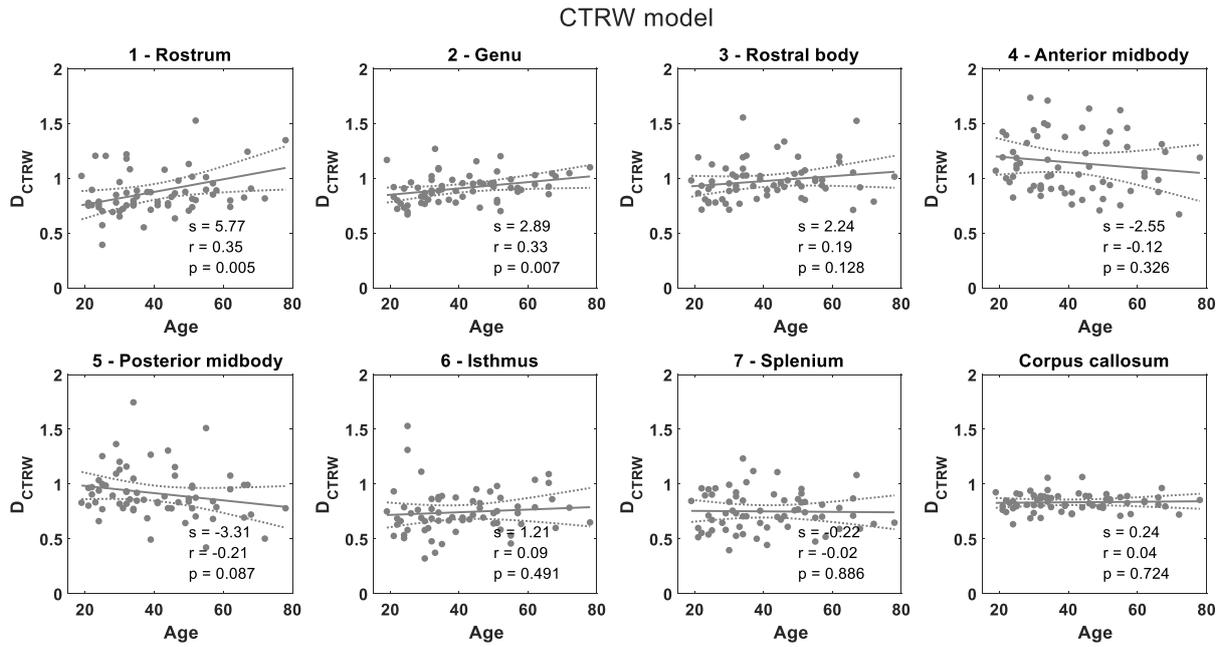

**Figure S9.** Association between age and $D_{CTRW}$ (in units of $10^{-3} \cdot mm^2/s$) from the sub-diffusion model within the corpus callosum and its seven subregions. Dots are the mean $D_{CTRW}$ in each ROI for each participant. Solid lines represent the linear trend between $D_{CTRW}$ and age. Dashed lines represent the 95% confidence interval. The slope of regression (s in units of $10^{-3} \cdot year^{-1}$), correlation coefficients (r) and the significance level of the linear relationship (p-values) are reported.

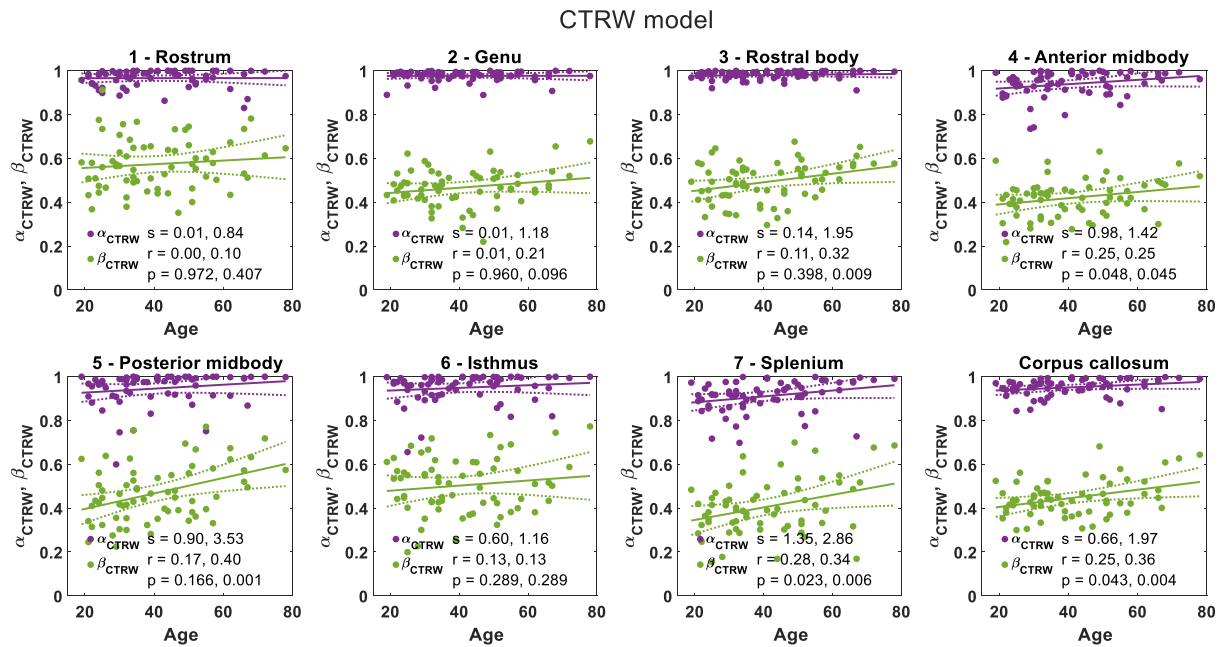

**Figure S10.** Association between age and $\alpha_{CTRW}$ and $\beta_{CTRW}$ from the continuous time random walk (CTRW) model within the corpus callosum and its seven subregions. Purple and green dots are the mean $\alpha_{CTRW}$ and $\beta_{CTRW}$

in each ROI, respectively, for each participant. Solid lines represent the linear trend between $\alpha_{CTRW}$ and $\beta_{CTRW}$ and age. Dashed lines represent the 95% confidence interval. The slope of regression (s in units of $10^{-3} \cdot year^{-1}$), correlation coefficients (r) and the significance level of the linear relationship (p-values) are reported. The two values in each measurement are for $\alpha_{CTRW}$ and $\beta_{CTRW}$, respectively.

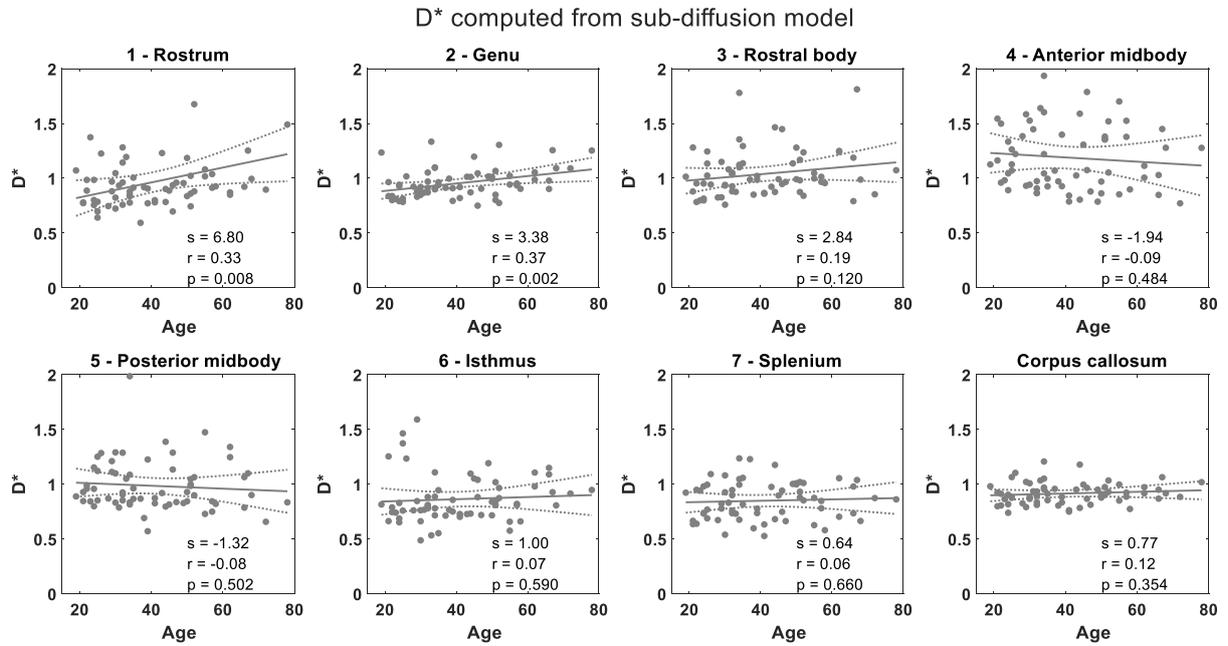

**Figure S11**. Association between age and $D^*$ (in units of $10^{-3} \cdot mm^2/s$) computed based on the sub-diffusion model parameters (refer to Eq.(16)) within the corpus callosum and its seven subregions. Dots are the mean $D^*$ in each ROI for each participant. Solid lines represent the linear trend between $D^*$ and age. Dashed lines represent the 95% confidence interval. The slope of regression (s in units of $10^{-3} \cdot year^{-1}$), correlation coefficients (r) and the significance level of the linear relationship (p-values) are reported.

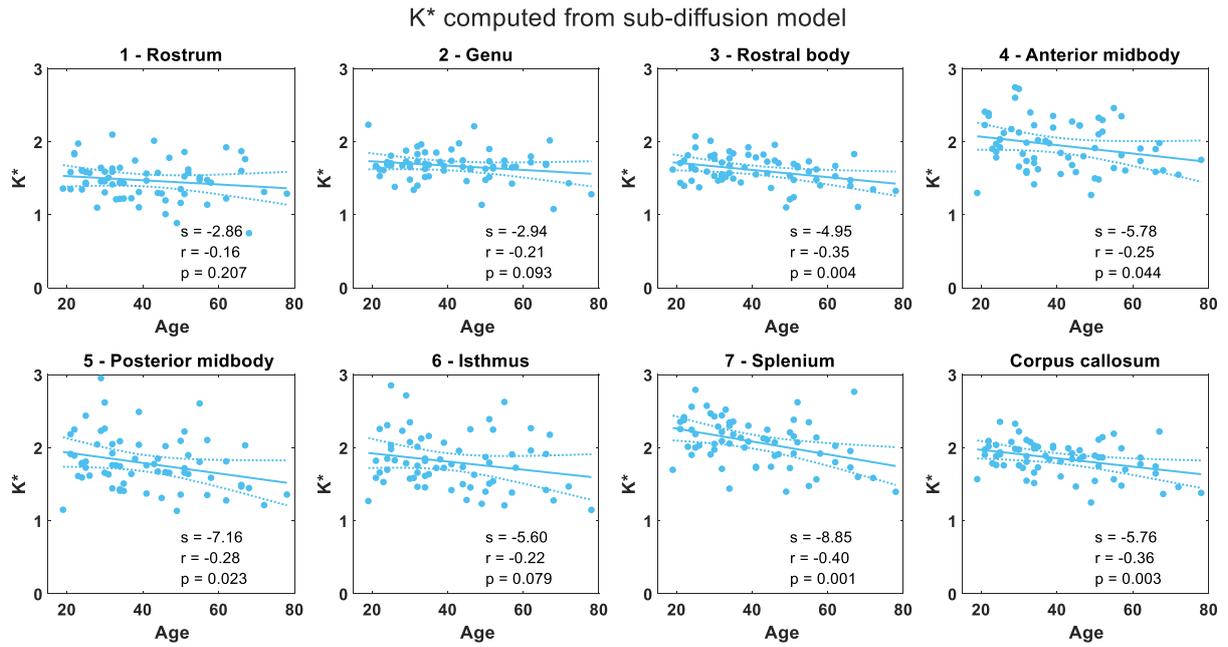

**Figure S12**. Association between age and $K^*$ computed based on the sub-diffusion model parameter (refer to Eq.(18)) within the corpus callosum and its seven subregions. Dots are the mean $K^*$ in each ROI for each participant. Solid lines represent the linear trend between $K^*$ and age. Dashed lines represent the 95% confidence interval. The slope of regression (s in units of $10^{-3} \cdot year^{-1}$), correlation coefficients (r) and the significance level of the linear relationship (p-values) are reported.

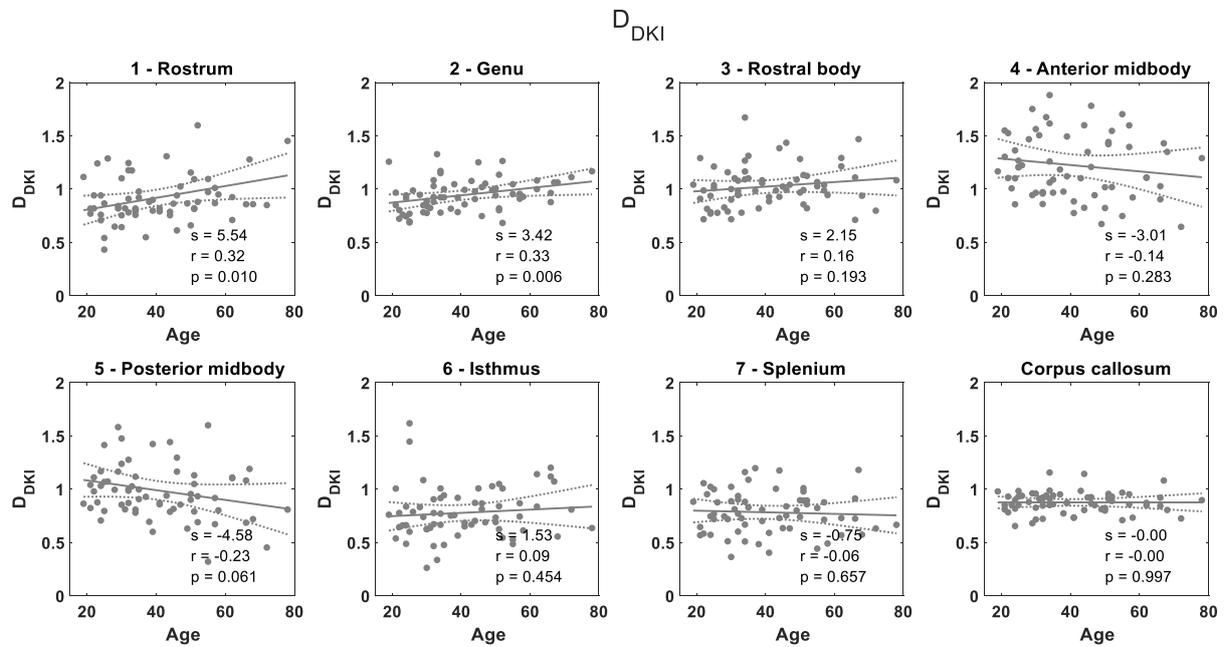

**Figure S13**. Association between age and $D_{DKI}$ (in units of $10^{-3} \cdot mm^2/s$) computed based on the sub-diffusion model parameters (refer to Eq.(16)) within the corpus callosum and its seven subregions. Dots are the mean $D_{DKI}$

in each ROI for each participant. Solid lines represent the linear trend between $D_{DKI}$ and age. Dashed lines represent the 95% confidence interval. The slope of regression (s in units of $10^{-3} \cdot year^{-1}$), correlation coefficients (r) and the significance level of the linear relationship (p-values) are reported.

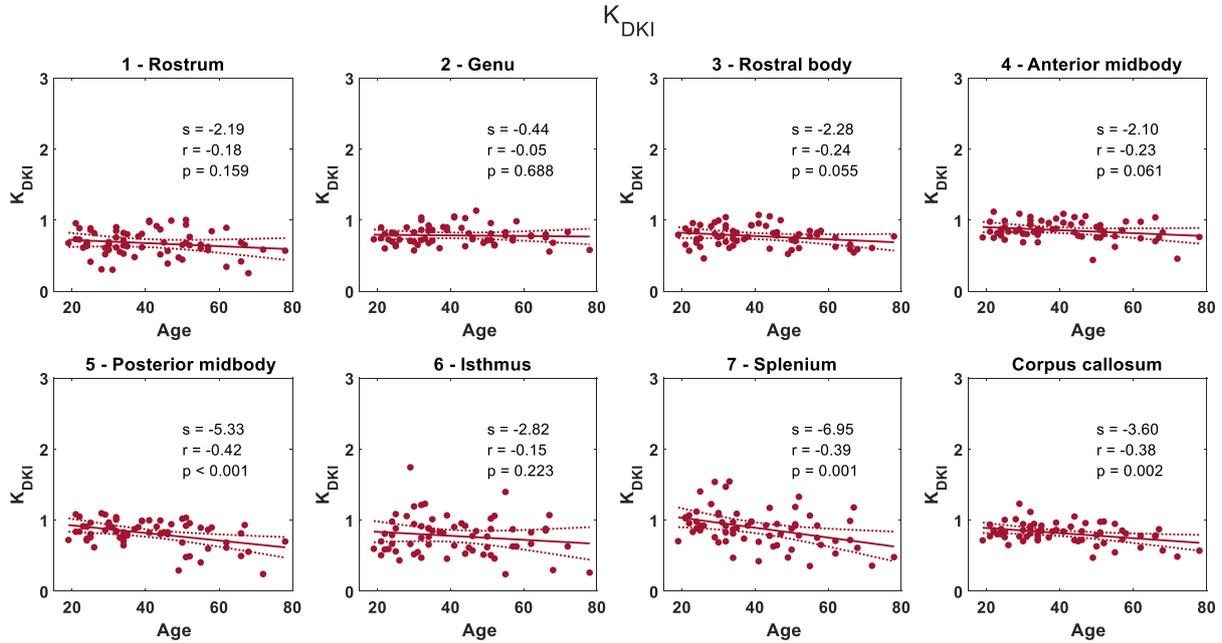

**Figure S14**. Association between age and $K_{DKI}$ computed based on the sub-diffusion model parameter (refer to Eq.(18)) within the corpus callosum and its seven subregions. Dots are the mean $K_{DKI}$ in each ROI for each participant. Solid lines represent the linear trend between $K_{DKI}$ and age. Dashed lines represent the 95% confidence interval. The slope of regression (s in units of $10^{-3} \cdot year^{-1}$), correlation coefficients (r) and the significance level of the linear relationship (p-values) are reported.